\documentclass[journal,draftclsnofoot,onecolumn,12pt,twoside]{IEEEtran}
\IEEEoverridecommandlockouts
\usepackage{amsmath}
\usepackage{amsfonts}
\usepackage{amssymb}
\usepackage{graphicx}
\usepackage{epstopdf}
\usepackage{caption}
\usepackage{subcaption}
\usepackage{cite}
\usepackage{algorithmicx}
\usepackage[ruled]{algorithm}
\usepackage{algpseudocode}
\usepackage{blindtext}
\usepackage{enumitem}
\usepackage{pdfpages}
\usepackage{bibentry}
\usepackage{amsthm}
\usepackage{bbm}

\begin{document}

\renewcommand{\baselinestretch}{1.33}

\title{Individual Preference Aware Caching Policy Design in Wireless D2D Networks}

\author{\IEEEauthorblockN{Ming-Chun Lee, \textit{Student Member, IEEE}, and Andreas F. Molisch, \textit{Fellow, IEEE}}
\IEEEauthorblockA{Ming Hsieh Department of Electrical and Computer Engineering\\
University of Southern California\\
Los Angeles, CA, USA\\
Email: mingchul@usc.edu, molisch@usc.edu}
\thanks{This work was supported in part by the National Science
Foundation (NSF). Part of this work has been presented at the 2017 IEEE Global Communications Conference \cite{Lee:Indi_op_EE}.}}


%Telephone: (800) 555--1212, Fax: (888) 555--1212}


\maketitle
\vspace{-1.8cm}
\begin{abstract}
%\boldmath
Cache-aided wireless device-to-device (D2D) networks allow significant throughput increase, depending on the concentration of the popularity distribution of files. Many studies assume that all users have the same preference distribution; however, this may not be true in practice. This work investigates whether and how the information about {\em individual} preferences can benefit cache-aided D2D networks. We examine a clustered network and derive a network utility that considers both the user distribution and channel fading effects into the analysis. We also formulate a utility maximization problem for designing caching policies. This maximization problem can be applied to optimize several important quantities, including throughput, energy efficiency (EE), cost, and hit-rate, and to solve different tradeoff problems. We provide a general approach that can solve the proposed problem under the assumption that users coordinate, then prove that the proposed approach can obtain the stationary point under a mild assumption. Using simulations of practical setups, we show that performance can improve significantly with proper exploitation of individual preferences. We also show that different types of tradeoffs exist between different performance metrics and that they can be managed through caching policy and cooperation distance designs. 
\end{abstract}

% \fi
\IEEEpeerreviewmaketitle

\vspace{-0.5cm}
\section{Introduction}
Over the years, wireless data traffic has rapidly increased, consequently straining wireless networks. This trend in data traffic is expected to continue over the next several years \cite{Cisco:5G}. Among all the wireless applications available today, on-demand video accounts for the largest portion of this traffic. Thus, finding an efficient approach to support this application is a paramount issue for modern wireless communication systems. Conventional approaches for increasing throughput, such as cell densification, installing large-scale antenna systems, and adopting millimeter-wave communications \cite{Andrews:5G}, are deemed insufficient, as such methods would entail high cost when investing in physical resources. Different from those approaches that tend to improve wireless networks without regard to the type of data to be transmitted, video caching at the wireless edge exploits the unique video accessing behavior of typical consumers and cheap storage resources to trade memory for bandwidth. In essence, different users cache different popular video files on their devices; a file request can then be satisfied either from a user's own cache or through D2D communication with a nearby user that has stored the requested file. The potential of D2D-based video caching has been widely discussed in recent years \cite{MeaPopDist:Lee,Ji:Th_Out_toff,Liu:Caching,Molisch:CachingEli,Gol:femtocaching,Gol:Dcache1,Ji:Dcache2,Ji:Dcache}. Accordingly, previously published papers have demonstrated, either in theory or in practice, that using wireless video caching with D2D communications can significantly improve throughput \cite{Ji:Dcache2,Ji:Th_Out_toff,Ji:Dcache,MeaPopDist:Lee}.

\subsection{Literature Review and Motivations}

Cache-aided D2D has demonstrated the ability to significantly improve network performance without the need for newly installed infrastructure and complicated coding.\footnote{Concentrated popularity distribution of video files can also be exploited in other ways, e.g., femtocaching \cite{Shanmugam:fcache,Chen:Coop_cach,Blaszczyszyn:fcache} and coded multicasting \cite{Maddah-Ali:NCCache,Ji:Dcache2}; those approaches are outside the scope of this paper.} Thus, numerous papers have been published on this topic. To understand cache-aided D2D from a theoretical point of view, \cite{Ji:Th_Out_toff,Ji:Dcache2,MeaPopDist:Lee} investigated the throughput-outage tradeoff. On the other hand, different performance metrics, including hit-rate (file outage) \cite{Chen:Dcache,Ji:Th_Out_toff}, throughput \cite{Ji:Th_Out_toff,Lee:caching,Chen:D2D_Coop}, energy efficiency (EE) \cite{EE:Chen,Lee:caching}, and latency \cite{Delay:Li,Gol:Dcache1} have also been investigated in order to improve the network from different aspects. Ref. \cite{Lee:caching} particularly studied the optimizations of throughput, EE, and their tradeoff by jointly considering the effect of the cooperation distance and caching policy of D2D users. Furthermore, \cite{EE:Chen} particularly focused on battery life when proposing an energy-efficient caching policy design. 

Different mathematical techniques and deployment scenarios were considered for cache-aided D2D. In \cite{Malak:Dcache} and \cite{Dhillon:StoCache}, stochastic geometry was exploited to analyze networks and to propose caching policy designs. Considering MIMO systems, scaling laws of throughput were discussed in \cite{Guo:Throughput} and \cite{Liu:Throughput}. In \cite{Wang:mobil_cach}, mobility was leveraged to enhance network performance. {In \cite{giatsoglou2017d2d}, a randomized caching policy with a special structure for helping content delivery through D2D links is proposed and analyzed in the systems that consider millimeter-wave communications for D2D links.} In \cite{lee2020dynamic}, a cache replacement approach was proposed to accommodate the environmental dynamics. Cache-aided D2D has been a subject of great interest to many researchers, and hundreds of related papers have accordingly been published on this topic. Hence, the above literature review, by necessity, cites only a sample of papers and topics. {To complement our literature review, we refer to several recent survey papers \cite{li2018survey,ahmed2019video,mehrabi2019device,prerna2020device}.}

Most of the existing papers for cache-aided D2D networks consider a homogeneous preference model, which assumes all users have the same file preference. In other words, each user requests files independently and randomly according to the {\em same} popularity distribution. However, this model is at best an approximation, because different users indeed have different tastes and preferences. Such heterogeneity in preferences of users has been observed in \cite{Karamshuk:PreCache} and has also been modeled in recent works \cite{Chen:D2D_indi,Lee:Indi_pre_model,Lee:Indi_pre_model_ToN}. Furthermore, based on real-world data, results in \cite{Karamshuk:PreCache} have shown that leveraging individual (user) preferences indeed can improve network performance. Thus, these abovementioned observations clarify that cache-aided D2D networks can be further improved by using a heterogeneous model, instead of a homogeneous model, in the network design. This is because designs that are based on the latter model are restricted, as it does not consider individual user preferences.

Researchers have recently begun to consider individual preferences in their analysis, and studies have accordingly shown that it is possible to use this information to improve the performance of wireless caching networks \cite{Zhang2018gam,Zhang:D2D_Indi,Chen:D2D_indi,Delay:Li,Liu:BS_Indi,Guo:PreCache_2,Karamshuk:PreCache,Pan:PreCache}.\footnote{The conference version of this paper \cite{Lee:Indi_op_EE} is one of the earliest studies that took individual preferences into consideration.} In \cite{Karamshuk:PreCache}, individual preferences were studied, and a machine-learning approach was used to learn the user's preferences and accordingly decide which video should be preloaded onto a local device cache based on the preferences of that particular user. While this kind of approach (also known as the ``Netflix challenge'') is very important for recommendation systems and preloading on individual devices, it is not applicable to cache-aided D2D networks. In \cite{Liu:BS_Indi}, an individual preference-aware weighted sum utility of users was formulated and optimized when the files were being cached at the BSs. Meanwhile, \cite{Guo:PreCache_2} designed a caching policy by assuming that users in different groups have different file preferences; the goal then is to maximize the successful file discovery probability of different groups without taking possible interference into account. In \cite{Pan:PreCache}, a content push strategy was designed to maximize the D2D offloading gain for a particular demand realization by jointly considering the influences of user preference and sharing willingness. In \cite{Chen:D2D_indi}, user preferences were used to maximize the offloading probability without accounting for the details of the physical layer. Using individual preference and user similarity, \cite{Zhang:D2D_Indi} proposed a caching content allocation approach to maximize a specifically defined utility. While \cite{Delay:Li} focused on estimating individual preferences using a learning-based algorithm, the study provided a caching policy that exploits the estimated preferences in order to minimize the average delay of D2D networks. Lastly, \cite{Zhang2018gam} proposed a game theoretical-cooperative caching design by assuming that users know exactly what files they want to request.

Despite this progress, the understanding of how individual preferences can be used to improve cache-aided D2D networks is still far from complete. It is still unclear whether integrating individual preferences into the design can improve network performance significantly. Moreover, the interplay between and among the different performance metrics, e.g., throughput, EE, and hit-rate, and the corresponding tradeoff that results from these interactions are still subject to further studies. Most importantly, the existing papers do not provide sufficient evaluations based on real-world data. Accordingly, our paper aims to address these issues.

\subsection{Contributions}
In this paper, we consider a BS-assisted cache-aided D2D network, where users can obtain the desired files from the BS, caches of neighboring users via D2D links, and their local caches. We assume that users have different preferences and caching policies; thus, our goal is to maximize network utility by designing individual preference-aware caching policies for users. We analyze the network based on the clustering and random-push scheduling \cite{Lee:caching} and then propose a non-convex utility maximization problem formulation. We then show that our proposed utility maximization problem can be applied to solve different practical problems, e.g., throughput, hit-rate, and EE optimization problems, as well as several tradeoff problems. Hence, it is sufficient to investigate a general solution approach for the proposed utility maximization problem. In addition, we discuss how the proposed utility maximization problem can be used in scenarios with different fading and user distributions.

We assume that users perfectly know the individual preferences of all the other users in the network and that they are allowed to coordinate with one another. With these considerations, we solve the utility maximization problem and obtain the users' caching policies. The idea of the proposed approach is to optimize individually and iteratively the users' caching policies until convergence. We show that this method is fairly simple to use, improves at each iteration, and converges to a stationary point under a mild assumption. We then evaluate the proposed caching policies in networks under realistic setups; in particular, we adopt the practical individual preference generator proposed in \cite{Lee:Indi_pre_model_ToN} based on extensive real-world data. 

Our results show that network performance can significantly improve when the information on individual preferences is properly exploited. We also compare the performances of those networks that optimize throughput, EE, and hit-rate using the proposed utility maximization framework and investigate the influences of the cooperation range of the D2D network. The results indicate that there are tradeoffs between these important metrics, and we can manage the tradeoffs by properly exploiting our proposed framework. Finally, we show how the proposed designs can be used as good reference designs for obtaining effective designs in networks with complicated scheduling. We emphasize that, to the best of our knowledge, this is the first work that validates from different perspectives of the network the benefits of exploiting user preferences and that gains insights through simulations based on real-world data. To sum up, our paper makes the following contributions:
\begin{itemize}
\item We formulate a utility maximization problem by considering individual preferences in the analysis. Caching policies that optimize several practically important metrics, e.g., throughput, EE, and hit-rate, and their tradeoffs, can be obtained by solving the problem. We then propose a general low-complexity approach for solving the utility maximization problem, and prove that the solution approach improves at each iteration and then converges to a stationary point.
\item Considering the realistic setup based on extensive real-world data, we conduct comprehensive simulations to show the benefits of exploiting individual preferences and to demonstrate tradeoffs between different performance metrics.
\end{itemize}

\subsection{Organization of This Paper}

The remainder of the paper is organized as follows. In Sec. II, we present the network and individual preference models. We then analyze the networks and formulate the utility maximization problem in Sec. III. Also in this section, we relate the proposed problem to different practical problems and show the effects of fading and user distributions. The caching policy design is proposed and studied in Sec. IV. Extensive simulation results are provided in Sec. V. We conclude the paper in Sec. VI. The proofs and detailed derivations are relegated to the Appendices.

\section{Network and Individual Preference Models}
We consider a BS-assisted cache-aided wireless D2D network, where the BS helps file delivery and makes scheduling decisions. The users can obtain the desired files by retrieving them from their own caches, D2D communications, or BS links. The file library consists of $M$ files, and for simplicity, we assume that all files have the same size.\footnote{This paper generally focuses on understanding the impact of individual preferences on network performance and the tradeoffs among different performance metrics. Thus, an investigation of how to deal with heterogeneous file sizes is beyond our scope. However, based on our numerical investigations (omitted for brevity), a performance evaluation using designs with equal filesize assumption could be representative of an evaluation without equal filesize assumption.} Each user is able to cache $S$ files in its storage. Besides, to have a nontrivial case, we require $S<M$. Furthermore, in most practical situations, $S<<M$ will hold. Users can be either active or inactive. An active user is a user that places a request that needs to be satisfied and participates in the D2D cooperation, i.e., the user sends files to other users that request them. On the other hand, an inactive user is a user that does not place its own request but still participates in the D2D cooperation.

\begin{figure}
\center
\includegraphics[width=0.6\textwidth]{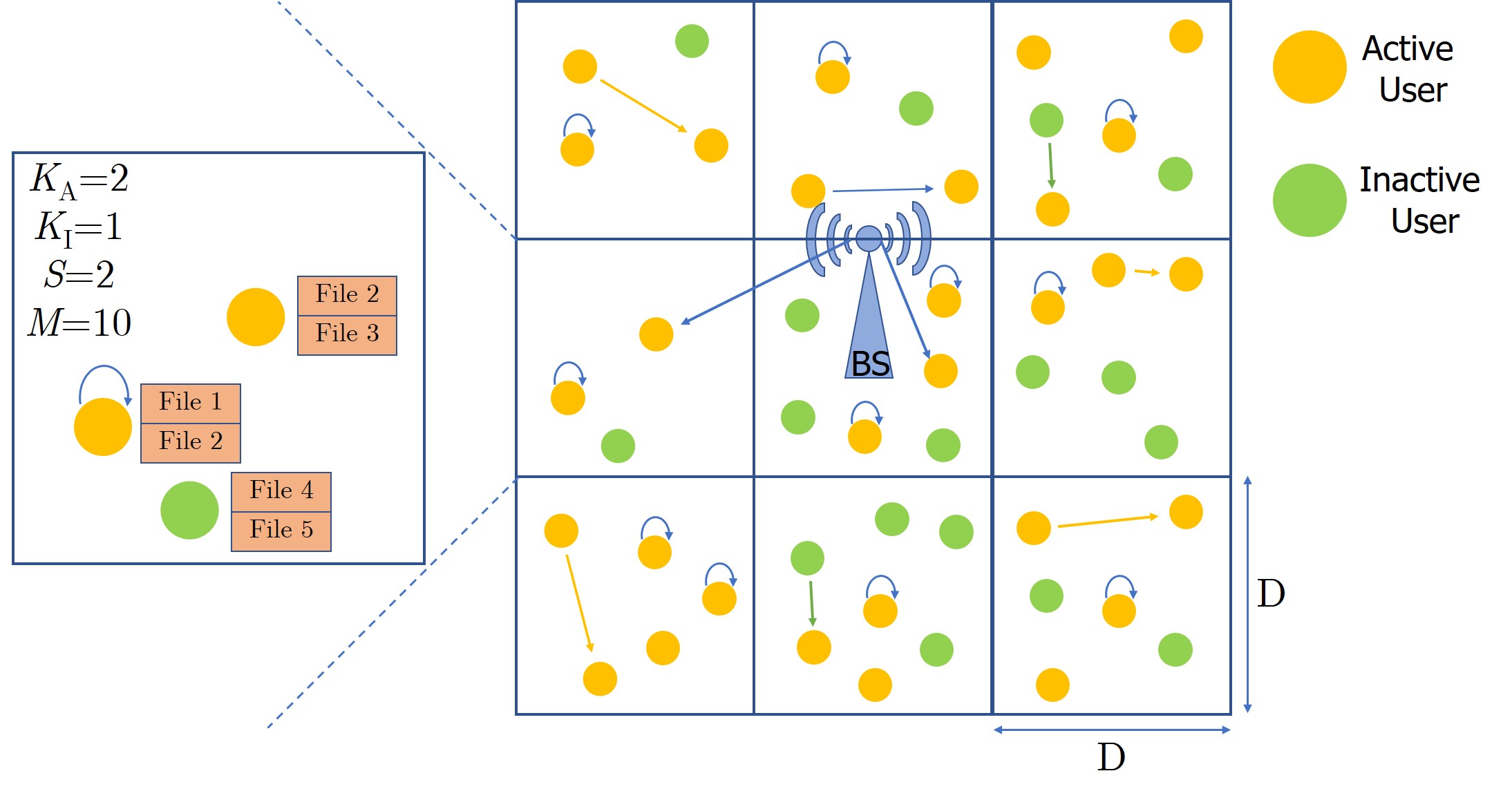}
\vspace{-15pt}
\caption{This figure shows an example of the network model. In the left-middle cluster, we have $K_{\text{A}}=2$, $K_{\text{I}}=1$, $S=2$, and $M=10$.}
\vspace{-25pt}
\label{fg:Fig_sys}
\end{figure}

We consider a widely used clustering network model \cite{Ji:Th_Out_toff,Chen:D2D_Coop,Lee:caching,Guo:Throughput,Liu:Throughput}. In the model, there is a square cell with a BS at the center point, and the cell is divided into equal-sized square clusters with side length $D$. The users are allowed to cooperate via D2D communications only with users in the same cluster. The ``cooperation distance'' or ``cluster size'' we henceforth reference thus corresponds to the dimension $D$ of such a cluster, not the cell radius of the BS. We assume that there is no interference between users of different clusters; this can be achieved by letting different clusters use different time/frequency resources with ``spatial reuse.'' We will in the following focus on a single cluster. Nevertheless, our results can easily be extended to multi-cluster scenarios. We denote the number of active users in a cluster as $K_{\text{A}}$ and the number of inactive users in a cluster as $K_{\text{I}}$. The total number of users in a cluster is then $K=K_{\text{A}}+K_{\text{I}}$. {The described model is shown in Fig. \ref{fg:Fig_sys}.}

We consider serving users via ``random-push'' scheduling \cite{Lee:caching}, which functions as follows. For a cluster, the BS first randomly selects an active user without knowing whether its request can be satisfied by the user's own (local) cache. If the selected user can obtain the desired file from the local cache, i.e., the desired file is actually cached by the selected user, then the user request is satisfied immediately. Otherwise, the BS checks whether the other users in the D2D network store the desired content and whether the channel quality between the selected user and the other users storing the desired file is larger than the minimum requirement (in terms of capacity) for a D2D transmission. If yes, then a D2D link is used to transmit the desired content; otherwise, the user needs to use the BS link to access the content. We assume that the BS has an unlimited bandwidth backhaul to repositories that store all files in the library. This guarantees that any request from a selected user can always be satisfied, albeit at a potentially high cost. After scheduling for the selected user, the remaining active users check whether the files in their local caches can satisfy their requests. If yes, then their requests are satisfied. Clearly, such scheduling can guarantee that at least one user is served and all users are scheduled fairly in the sense that every user is selected with equal probability by the BS for service. We note that there exist scheduling approaches with better overall throughput, e.g., priority-push scheduling \cite{Lee:caching} and dynamic link scheduling \cite{Zhang:D2D_Schedule}. However, using them has other drawbacks, such as unfairness and high complexity. Most importantly, it is very challenging to obtain tractable formulations for these advanced scheduling schemes \cite{Lee:caching,Zhang:D2D_Schedule}. On the other hand, random-push scheduling leads to tractable expressions for different critical metrics and is easy to implement, thus serving as a good reference system. It should be noted that the definitions of user distribution and channel model can influence the scheduling behaviors. We will discuss them later in Sec. III.D.

We represent individual preferences for requesting video files as probabilities. We denote the request probability of user $k$ for file $m$, i.e., the probability that user $k$ wants file $m$ in the future, as $a_m^k$, where $0\leq a_m^k\leq 1,\forall m,k$ and $\sum_{m=1}^M a_m^k=1,\forall k$. Note that the preferences of inactive users are actually not used in the proposed problem formulation and solution approach later; their modeling of preferences is for the purpose of consistency. Different users can have different caching policies. As such, we denote $b_m^k$ as the probability that user $k$ would cache file $m$, whereas the caching policy of user $k$ is described by $\lbrace b_m^k \rbrace_1^M$, where $0\leq b_m^k\leq 1,\forall m,k,$ and $\sum_{m=1}^M b_m^k\leq S,\forall k$. An implementation of this probabilistic caching policy can be found in \cite{Blaszczyszyn:fcache}. Note that the caching policy becomes deterministic when considering the limiting case that $b_k^m$ becomes 1 or 0. This is useful in the situation where the central controller knows a priori which users are going to be in a cluster, so that the caching policy can avoid detrimental file overlap (compare \cite{Gol:Dcache1}). Such a deterministic predictability of user location occurs in a place where the same people are in geographical proximity every day, e.g., in an office scenario.

We assume that users perfectly know the individual preferences of the other users in the same cluster. In other words, users know $a_m^k,\forall m,k$. We also assume that users can coordinate with one another to design their caching policies for a common goal, e.g., maximizing network throughput. The users then coordinate such that they cache files by fully considering the caching policies and preferences of other users.\footnote{A common approach used to incentivize users to coordinate is through payment by the network operator. Alternatively, since each user generally benefits from D2D communications, a token-based approach similar to the traditional file-sharing networks can ensure that specific users do not exploit the system without contributing to it. Generally, the topic of giving incentives is an important one for D2D networks. However, this is already beyond the scope of this paper, albeit it is still considered an important future direction of this paper.} Accordingly, this approach suits scenarios where user locations are deterministic, e.g., office scenarios. Note that the information exchange and the coordination between users are assumed to be handled centrally by the BS. Moreover, it will be shown later that although users are assumed to coordinate, the algorithm does not necessarily need to be operated in a centralized manner. On the contrary, since users have a common goal and know the individual preferences of other users perfectly, each user can independently implement the same caching policy design algorithm and accordingly obtain the same coordinated design that gives the policies of all users. Consequently, each user can extract its own caching policy and then implement independently.

In this work, users can access the desired files from their local caches, from the caches of other users, and from the BS. We consider different utilities when different types of approaches are used. The utility of accessing a file via a BS link is then denoted as $U_{\text{B}}$, the utility of accessing a file via a D2D link as $U_{\text{D}}$, and the utility of accessing a file via the user's own cache as $U_{\text{S}}$. Although we consider all users to have the same utility, the extension to the case that different users have different utilities is straightforward. Besides, the utility can be set differently for different practical purposes, such as throughput and EE maximization, etc. However, we will generally assume that $U_{\text{B}}\leq U_{\text{D}}\leq U_{\text{S}}$, which implies that using self-access is superior to using a D2D link, and using a D2D link is superior to using a BS link. We will discuss this more thoroughly in Sec. III.C. Table \ref{tb:1} summarizes the notations frequently used in this paper.
\begin{table}
\caption{Summary of Frequently Used Notations}
\centering
\begin{tabular}{| c | c | c | c |}
\hline
Notations & Descriptions \\
\hline
$M$; $S$; $D$  & Number of files in the library; number of files that can be stored by a user; cluster size  \\
\hline
$K_{\text{a}}$; $K_{\text{i}}$; $K$ & Number of active users; of inactive users; of total users in a cluster \\
\hline
$\mathcal{U}_{\text{A}}$; $\mathcal{U}_{\text{I}}$; $\mathcal{U}$ & Index set of active users ; of inactive users; of all users \\
\hline
$a_m^k$; $b_m^k$ & Request probability of user $k$ for file $m$; caching probability of user $k$ for file $m$  \\
\hline
$L_{k,l}$; $w_k$ & Probability of a successful D2D link between users $k$ and $l$; weight of user $k$ \\
\hline
$U_{\text{B}}$; $U_{\text{D}}$; $U_{\text{S}}$ & Utility of using BS link; of using D2D link; of using self-caching \\
\hline
$T_{\text{B}}$; $T_{\text{D}}$; $T_{\text{S}}$ & Throughput of using BS link; of using D2D link; of using self-caching \\
\hline
$C_{\text{B}}$; $C_{\text{D}}$; $C_{\text{S}}$ & Cost of using BS link; of using D2D link; of using self-caching \\
\hline
$U_{\text{net}}$; $T_{\text{net}}$; $C_{\text{net}}$; $H_{\text{net}}$; $\text{EE}_{\text{net}}$ & Utility; throughput; cost; energy efficiency; hit-rate of the network \\ 
\hline
$P_{\text{B}}^k$; $P_{\text{D}}^k$; $P_{\text{S}}^k$ & Elementary access probabilities: refer to the clear definition of (1); of (3); of (2) \\
\hline
\end{tabular}
\label{tb:1}
\end{table}

\section{Caching Policy Design Problem}
Our goal here is to design caching policies that optimize network utility by using information about individual preferences. In this section, we first derive the access probabilities of different accessing approaches for a user. Based on the results, we then formulate the caching policy design problem that we aim for. To clarify the usefulness of the proposed network utility maximization problem, we then show how it can be used to solve various practical problems. Finally, we discuss how the proposed network utility can accommodate different scenarios with different fading and user distributions.

\subsection{Fundamental Access Probability}
Consider the system model in Sec. II. We denote $\mathcal{U}_A$ as the index set of active users and $\mathcal{U}_I$ as the index set of inactive users, and $\mathcal{U}= \mathcal{U}_A \bigcup \mathcal{U}_I$. We denote the channel between user $k$ and user $l$ as $h_{k,l}$ and the corresponding signal-to-noise ratio (SNR) as $\text{SNR}_{k,l}$. We let $C$ be the minimal capacity requirement for establishing a D2D link. When user $k$ is selected, the probability that user $k$ accesses the desired file through a BS link is expressed as
\begin{equation}
\label{eq:BS_rate}
P_{\text{B}}^k=\sum_{m=1}^M a_m^k\left[\prod_{l\in \mathcal{U}}\left(1-b_m^l\mathbf{1}_{\lbrace h_{k,l},C\rbrace}\right)\right],
\end{equation}
where $\mathbf{1}_{\lbrace h_{k,l},C\rbrace}=1$ if $\log_2(1+\text{SNR}_{k,l})> C$; otherwise $\mathbf{1}_{\lbrace h_{k,l},C\rbrace}=0$. Note that $\displaystyle{\prod_{l\in \mathcal{U}}\left(1-b_m^l\mathbf{1}_{\lbrace h_{k,l},C\rbrace}\right)}$ is the probability that file $m$ can be obtained only via a BS link, and $\displaystyle{a_m^k\prod_{l\in \mathcal{U}}\left(1-b_m^l\mathbf{1}_{\lbrace h_{k,l},C\rbrace}\right)}$ is the probability that the user wants file $m$ but file $m$ can be obtained only via a BS link. We define the self-access probability, i.e., the probability that user $k$ can obtain the desired file from its own cache, as
\begin{equation}
\label{eq:self_rate}
P_{\text{S}}^k=\sum_{m=1}^M a_m^k b_m^k.
\end{equation}
By using $P_{\text{B}}^k$ and $P_{\text{S}}^k$, the probability that user $k$ obtains the desired file via a D2D link is
\begin{equation}
\label{eq:D2D_rate}
P_{\text{D}}^k=1-P_{\text{S}}^k-P_{\text{B}}^k= 1-\sum_{m=1}^M a_m^k\left[\prod_{l\in \mathcal{U}}\left(1-b_m^l\mathbf{1}_{\lbrace h_{k,l},C\rbrace}\right)\right]-\sum_{m=1}^M a_m^k b_m^k.
\end{equation}

\subsection{Utility Maximization Problem Formulation}
Now, we derive the expected utility of the network. We assume that for any user $k$, the channel gains of all possible associated D2D links, i.e., $\mathbf{1}_{\lbrace h_{k,l},C\rbrace},\forall  l,$ are independent (see use cases in Sec. III.D). Using the results in Sec. III.A, the utility of the selected user $k$ when user $k$ is selected by the BS is expressed as
\begin{equation}
\label{eq:EE_k}
U_k=U_{\text{D}}\cdot P_{\text{D}}^k+U_{\text{B}}\cdot P_{\text{B}}^k + U_{\text{S}}\cdot P_{\text{S}}^k.
\end{equation}
We denote weights $w_1,w_2,...,w_{K_{\text{A}}}$ as the weighting on different users, which indicates the relative priority of users. Since users are randomly selected by the BS, the expected utility contributed by the selected users is
\begin{equation}
\label{eq:EE}
\begin{aligned}
U &=\sum_{k\in\mathcal{U}_A} w_k \mathbb{E}\lbrace U_{\text{D}}\cdot P_{\text{D}}^k+U_{\text{B}}\cdot P_{\text{B}}^k + U_{\text{S}}\cdot P_{\text{S}}^k\rbrace=\sum_{k\in\mathcal{U}_A} \frac{w_k}{K_{\text{A}}}\left[ U_{\text{D}}\cdot P_{\text{D}}^k+U_{\text{B}}\cdot P_{\text{B}}^k + U_{\text{S}}\cdot P_{\text{S}}^k\right]
\end{aligned}
\end{equation}
The users not selected by the BS can still check whether their desired files are cached in their local caches. As such, we can obtain additional utilities from the users' ability to satisfy their own requests. Thus, the expected utility of the network is
\begin{equation}
\begin{aligned}\label{Ut_reformulation_sum}
U_{\text{net}}&=U+U_{\text{local}}=U+U_{\text{S}}\cdot\frac{1}{K_{\text{A}}}\cdot\sum_{k\in\mathcal{U}_A}\sum_{l\in\mathcal{U}_A,l\neq k} \sum_{m=1}^{M}w_la_m^lb_m^l\\
&=\sum_{k\in\mathcal{U}_A} \frac{w_k U_{\text{D}}}{K_{\text{A}}}+(U_{\text{B}}-U_{\text{D}})\sum_{m=1}^M S_m +(K_{\text{A}}U_{\text{S}}-U_{\text{D}})\sum_{m=1}^M\sum_{k\in\mathcal{U}_A} \frac{w_ka_m^kb_m^k}{K_{\text{A}}},
\end{aligned}
\end{equation}
where $S_m=\sum_{k\in\mathcal{U}_A} \frac{w_ka_{m,k}}{K_{\text{A}}}\prod_{l\in \mathcal{U}}(1-b_m^lL_{k,l})$ and $L_{k,l}=\text{Pr}\left[\log_2(1+\text{SNR}_{k,l})> C\right]$. Note that the derivations for (\ref{Ut_reformulation_sum}) are shown in Appendix A. Moreover, the computation for $L_{k,l}$ will be discussed later in detail in Sec. III.D.

Using (\ref{Ut_reformulation_sum}), the caching policy design problem that maximizes the network utility is:
\begin{equation}\label{eq:optmization_Ut}
\begin{array}{cl}
\displaystyle{\max_{b_m^k,\forall k,m}} & U_{\text{net}}\\
\text{subject to} & \sum_{m=1}^M b_m^k\leq S,\forall k,\\
& 0\leq b_m^k\leq 1,\forall k,m.\\
\end{array}
\end{equation}
We then have the following proposition.

\textit{Proposition 1:} The optimal solution of (\ref{eq:optmization_Ut}) must be tight at the equality of the sum constraint, i.e., for the optimal solution $(b_m^{k})^*,\forall k,m$, we have
\begin{equation}
\sum_{m=1}^M (b_m^{k})^* = S,\forall k.
\end{equation}
\begin{proof}
By (\ref{Ut_reformulation_sum}), the first-order partial derivative of $U_{\text{net}}$ is:
\begin{equation}
\frac{\partial U_{\text{net}}}{\partial b_m^j}=-(U_{\text{B}}-U_{\text{D}})\sum_{k\in\mathcal{U}_A}\frac{w_ka_m^k}{K_{\text{A}}}L_{k,j}\prod_{l\in \mathcal{U},l\neq j}(1-b_m^lL_{k,l})+\mathbf{1}_{\lbrace j\in\mathcal{U}_{\text{A}}\rbrace}(K_{\text{A}}U_{\text{S}}-U_{\text{D}})\frac{w_ja_m^j}{K_{\text{A}}},\forall j,m,
\end{equation}
where $\mathbf{1}_{\lbrace j\in\mathcal{U}_{\text{A}}\rbrace}=1$ when $j\in\mathcal{U}_{\text{A}}$; otherwise $\mathbf{1}_{\lbrace j\in\mathcal{U}_{\text{A}}\rbrace}=0$. Since $U_{\text{B}}\leq U_{\text{D}}\leq U_{\text{S}}$, $0\leq L_{k,l}\leq 1,\forall k,l$, and $0\leq b_m^k\leq 1,\forall k,m$, we then have $\frac{\partial U_{\text{net}}}{\partial b_m^k}\geq 0,\forall k,m$. Therefore, $U_{\text{net}}$ is non-decreasing with respect to $b_m^k,\forall k,m$, which indicates that the optimal solution of (\ref{eq:optmization_Ut}) must be tight at the equality of the sum constraint.
\end{proof}

\subsection{Interpretations of the Utility Maximization Problem and Its Relationship to Practice}
In this subsection, we show how the utility maximization problem can be used in designing caching policies to solve various practical and important problems. In the following, we consider the equal-weight case, i.e., $w_1=w_2=...=w_{K_{\text{A}}}=1$, for notation convenience. The extension to other weights is straightforward.

\subsubsection{Throughput Maximization Problem} 
Consider $U_{\text{B}}=T_{\text{B}}$, $U_{\text{D}}=T_{\text{D}}$, $U_{\text{S}}=T_{\text{S}}$, and $T_{\text{B}}\leq T_{\text{D}}\leq T_{\text{S}}$, where $T_{\text{B}}$ is the throughput of a BS link, $T_{\text{D}}$ is the throughput of a D2D link, and $T_{\text{S}}$ is the throughput of self-access. The utility maximization problem then becomes the throughput maximization problem, in which the expected throughput is  
\begin{equation}
T_{\text{net}}=T_{\text{D}}+(T_{\text{B}}-T_{\text{D}})\sum_{m=1}^M S_m + (K_{\text{A}}T_{\text{S}}-T_{\text{D}})\sum_{m=1}^M\sum_{k\in \mathcal{U}_A}\frac{a_m^kb_m^k}{K_{\text{A}}}.
\end{equation}

\subsubsection{Cost/Power Minimization Problem} 
Let $U_{\text{B}}=-C_{\text{B}}$, $U_{\text{D}}=-C_{\text{D}}$, $U_{\text{S}}=-C_{\text{S}}$, and $C_{\text{B}}\geq C_{\text{D}}\geq C_{\text{S}}$, where $C_{\text{B}}$ is the cost of a BS link, $C_{\text{D}}$ is the cost of a D2D link, and $C_{\text{S}}$ is the cost of self-access. The problem can then be cast as the cost minimization problem, expressed as
\begin{equation}\label{eq:optmization_Cost}
\begin{array}{cc}
\displaystyle{\min_{b_m^k,\forall k,m}} & C_{\text{net}}=C_{\text{D}}+(C_{\text{B}}-C_{\text{D}})\sum_{m=1}^M S_m + (K_{\text{A}}C_{\text{S}}-C_{\text{D}})\sum_{m=1}^M\sum_{k\in \mathcal{U}_A} \frac{a_m^kb_m^k}{K_{\text{A}}}\\
\text{subject to} & \sum_{m=1}^M b_m^k\leq S,\forall k,0\leq b_m^k\leq 1,\forall k,m.\\
\end{array}
\end{equation}
If the power consumption is considered as cost, the problem is the power minimization problem.

\subsubsection{Hit-Rate Maximization Problem}
Let $U_{\text{B}}=0$, $U_{\text{D}}=1$, and $U_{\text{S}}=\frac{1}{K_{\text{A}}}$. The problem then is to maximize
\begin{equation}\label{Hit_reformulation_sum}
\begin{aligned}
&H_{\text{net}} =\sum_{k\in \mathcal{U}_A} \mathbb{E}\left[ P_{\text{D}}^k + P_{\text{S}}^k\right]=1-\sum_{m=1}^M S_m,
\end{aligned}
\end{equation}
which maximizes the file hit-rate of the network.

\subsubsection{Throughput--Cost Weighted Sum Problem}
To attain the desired tradeoff between the different metrics, a common approach is to maximize the weighted sum/difference of the different metrics \cite{MultiOpt:Ehrgott}. For example, considering the tradeoff between throughput and cost, we can maximize
\begin{equation}
w_{\text{T}}T_{\text{net}}-w_{\text{C}}C_{\text{net}},
\end{equation}
where $w_{T}\geq 0$ and $w_{C}\geq 0$. Such a weighted sum/difference problem is equivalent to the utility maximization problem, as we let $U_{\text{B}}=w_{\text{T}}T_{\text{B}}-w_{\text{C}}C_{\text{B}}$, $U_{\text{D}}=w_{\text{T}}T_{\text{D}}-w_{\text{C}}C_{\text{D}}$, and $U_{\text{S}}=w_{\text{T}}T_{\text{S}}-w_{\text{C}}C_{\text{S}}$. The same concept can also be used for the throughput--hit-rate tradeoff. Also, the same concept can be applied to tradeoff between more than two objectives.

\subsubsection{Efficiency Problem}
In some situations, we aim to maximize the efficiency, e.g., EE (bits/Joule). The following discussions then show that the efficiency maximization problem can be addressed by solving the weighted sum problem described in Sec. III.C.4.

We consider the EE maximization problem as an example. The same concept can be used for other problems. Suppose we aim to maximize EE, which is given as
\begin{equation}
\begin{aligned}\nonumber
\text{EE}_{\text{net}}&=\frac{\text{total number of bits transmitted}}{\text{total amount of energy consumed}}=\frac{\text{expected throughput of the network}}{\text{expected power consumed by the network}}=\frac{T_{\text{net}}}{C_{\text{net}}}.
\end{aligned}
\end{equation} 
Then, the EE maximization problem is:
\begin{equation}\label{eq:optmization_EE}
\begin{array}{cc}
\displaystyle{\max_{b_m^k,\forall k,m}} & \text{EE}_{\text{net}}=\frac{T_{\text{net}}}{C_{\text{net}}}=\frac{T_{\text{D}}+(T_{\text{B}}-T_{\text{D}})\sum_{m=1}^M S_m + (K_{\text{A}}T_{\text{S}}-T_{\text{D}})\sum_{m=1}^M\sum_{k\in \mathcal{U}_A}\frac{a_m^kb_m^k}{K_{\text{A}}}}{C_{\text{D}}+(C_{\text{B}}-C_{\text{D}})\sum_{m=1}^M S_m + (K_{\text{A}}C_{\text{S}}-C_{\text{D}})\sum_{m=1}^M\sum_{k\in \mathcal{U}_A}\frac{a_m^kb_m^k}{K_{\text{A}}}}\\
\text{subject to} & \sum_{m=1}^M b_m^k\leq S,\forall k,0\leq b_m^k\leq 1,\forall k,m.\\
\end{array}
\end{equation}
This problem is then equivalent to
\begin{equation}\label{eq:optmization_EE_2}
\begin{array}{cc}
\displaystyle{\max_{t,b_m^k,\forall k,m}} & t\\
\text{subject to} & \frac{T_{\text{net}}}{C_{\text{net}}}\geq t,\\
& \sum_{m=1}^M b_m^k\leq S,\forall k,0\leq b_m^k\leq 1,\forall k,m,\\
\end{array}
\end{equation}
Assuming that the optimal $t^*$ is known, then the problem in (\ref{eq:optmization_EE_2}) is equivalent to finding the optimal policy in
\begin{equation}\label{eq:optmization_EE_4}
\begin{array}{cl}
\displaystyle{\max_{b_m^k,\forall k,m}} & T_{\text{net}}-t^* C_{\text{net}} \\
\text{subject to} & \sum_{m=1}^M b_m^k\leq S,\forall k,0\leq b_m^k\leq 1,\forall k,m.\\
\end{array}
\end{equation}
In observing (\ref{eq:optmization_EE_4}), we see clearly that we have a weighted difference problem similar to that described in Sec. III.C.4, in which $w_{\text{T}}=1$ and $w_{\text{C}}=t^*$. Thus, it can be cast into the utility maximization framework. Also, the optimal policy should result in $T_{\text{net}}-t^* C_{\text{net}}=0$.

In general, we cannot know the optimal $t^*$ a priori; however, the aforementioned idea can still be used to solve the EE maximization problem. Suppose we have the same problem as that in (\ref{eq:optmization_EE_4}), but we now replace $t^*$ with $t$. Accordingly, we have the following interpretations: (i) if the solution results in a positive number, i.e., $T_{\text{net}}-t C_{\text{net}}> 0$, then our solution can provide an EE larger than $t$; (ii) if the solution gives $T_{\text{net}}-t C_{\text{net}}< 0$, then our solution provides an EE that is less than $t$, and $t$ is not achievable. As such, by adjusting $t$ based on the results and by solving the problem in (\ref{eq:optmization_EE_4}) using different $t$, we can keep optimizing $t$. Thus, we improve the solution. Finally, by carefully adjusting $t$ and by solving (\ref{eq:optmization_EE_4}) many times, we can maximize the EE. Since the utility maximization problem is non-convex, we might not find the best $t^*$ and the corresponding user caching policies. However, we can still obtain an effective solution by using the above approach. This technique is identical to that used for solving a quasi-convex problem \cite{CVX:Boyd}.

\subsection{Effects of the Statistics of Wireless Channels and User Distributions}
In (\ref{Ut_reformulation_sum}), the channel quality influences the expected utility via $L_{k,l}$. Thus, understanding the general expression of $L_{k,l}$ and its relationship to channel physics is important. In this section, we provide several useful expressions for $L_{k,l}$ and then discuss its relationship to the possible scenarios. Note that if $k=l$, then $L_{k,l}=L_{k,k}=1$. Therefore in the following, we consider $k\neq l$.

Let $d_{k,l}$ be the distance between user $k$ and user $l$.  The input--output relationship between users $k$ and $l$ then follows the general expression:
\begin{equation}\label{eq:channel_model}
y_l=\sqrt{\text{PG}(d_{k,l})s_{k,l}}h_{k,l}x_k + n_l,
\end{equation}
where $y_l$ is the received signal at user $l$; $x_k$ is the transmit signal from user $k$; $\text{PG}(d_{k,l})$ is the path gain effect (channel [power] gain averaged over small-scale and large-scale fading); $s_{k,l}$ is the shadowing power gain; $h_{k,l}$ is the small-scale fading amplitude; and $n_l$ is the Gaussian noise with power $\sigma_n^2$. Let $E_{\text{D}}$ be the transmission power of the D2D link. Using (\ref{eq:channel_model}), the received SNR for the D2D link between users $k$ and $l$ is $\text{SNR}_{k,l}=\frac{E_{\text{D}} \vert h_{k,l}\vert^2s_{k,l} \text{PG}(d_{k,l})}{\sigma_n^2}$, and therefore
\begin{equation}\label{eq:channel_qual_prob}
L_{k,l}=\text{Pr}\left[\vert h_{k,l}\vert^2s_{k,l}\text{PG}(d_{k,l})>\frac{\sigma_n^2(2^C-1)}{E_{\text{D}}}\right].
\end{equation}
We will show later some practical examples and then demonstrate how (\ref{eq:channel_qual_prob}) is computed using user and fading distributions. The extensions to other models are feasible by leveraging the existing results of fading \cite{WireCom:Molisch} and distance distributions \cite{Mathai:Dis_Dist}.

\subsubsection{Case 1: Systems with effective link quality control}
In modern wireless communication systems, approaches such as adaptive power control and frequency-and-antenna-diversity are used to combat fading effects in wireless channels. Thus, in systems with effective link quality control, we can assume that the D2D links between users in an area can be guaranteed, leading to $L_{k,l}=1,\forall k, l$. The exact distribution of users then becomes irrelevant in this case.

\subsubsection{Case 2: Systems with deterministic path-loss and shadow fading}
When users are less mobile or stationary, the joint effect of pathloss and shadow fading between users is deterministic. As a result, $s_{k,l}$ and $\text{PG}(d_{k,l})$ are the given constants and are based on the exact locations of users. In this case, we focus on characterizing the small-scale fading. Thus,
\begin{equation}
L_{k,l}=\text{Pr}\left[\vert h_{k,l}\vert^2 >\frac{\sigma_n^2(2^C-1)}{E_{\text{D}}s_{k,l}\text{PG}(d_{k,l})}\right],
\end{equation}
where the closed-form expressions are attainable for commonly used fading distribution. For example, let us consider a normalized Rayleigh fading whose average power is $1$; we have
\begin{equation}
L_{k,l}=\exp\left[\frac{-\sigma_n^2(2^C-1)}{E_{\text{D}}s_{k,l}\text{PG}(d_{k,l})}\right].
\end{equation}
Note that in this case, the distribution of users can be arbitrary but deterministic.

\subsubsection{Case 3: $K$ users uniformly distributed in a square with side length $D$ and with shadowing and small-scale fading}
Here, we use the lognormal shadowing and normalized Rayleigh fading as an example. According to results in \cite{Tseng:Dis_Dist}, the distance $d$ between two users independently and uniformly distributed over a square area with unit side length is described by the probability density function:
\begin{equation}
f_{\text{sq}}(d)=\left\{
\begin{aligned}
&2d(\pi+d^2-4d),&0\leq d\leq 1,\\
&2d(-2-d^2+4\sqrt{d^2-1}+2\sin^{-1}\frac{2-d^2}{d^2}),&1<d\leq\sqrt{2}.
\end{aligned}
\right.
\end{equation}
Then, when fixing the shadowing $s_{k,l}$, again using the property of Rayleigh fading, we have
\begin{equation}
\begin{aligned}\label{eq:aaa}
L_{k,l}(s_{k,l})&=\int_0^{\sqrt{2}D}\exp\left[- \frac{\sigma_n^2(2^C-1)}{E_{\text{D}}s_{k,l}\text{PG}(x)}\right]f[d=x])dx\\
&=\int_0^{\sqrt{2}}\exp\left[- \frac{\sigma_n^2(2^C-1)}{E_{\text{D}}s_{k,l}\text{PG}(Dx)}\right]f_{\text{sq}}(x))dx.
\end{aligned}
\end{equation}
Assume that the shadowing and small-scale fading effects of different links between different users are independent. We can then generalize (\ref{eq:aaa}) as
\begin{equation}
L_{k,l}=\int_0^{\sqrt{2}}\left[\int_{0}^{\infty}\exp\left[- \frac{\sigma_n^2(2^C-1)}{E_{\text{D}}s\text{PG}(Dx)}\right]f_{s_{k,l}}(s)ds\right]f_{\text{sq}}(x))dx,
\end{equation}
where $f_{s_{k,l}}(s)$ is the pdf of the shadowing effect for the channel link between user $k$ and $l$. Let the mean and standard deviation of the lognormal distribution be $u_{dB}$ and $\sigma_F$, respectively. We then obtain
\begin{equation}\label{eq:general_L}
L_{k,l}=\int_0^{\sqrt{2}}\left[\int_{0}^{\infty}\exp\left[- \frac{\sigma_n^2(2^C-1)}{E_{\text{D}}\text{PG}(Dx)s}\right]\frac{10/\log(10)}{s\sigma_F \sqrt{2\pi}}\exp\left(\frac{ -(10\log_{10}(s)-u_{dB})^2}{2\sigma_F^2}\right)ds\right]f_{\text{sq}}(x)dx.
\end{equation}
It should be noted that the inner integral of (\ref{eq:general_L}) is the complement of the channel outage when the joint effect of the fading is the Suzuki distribution \cite{WireCom:Molisch}.

\section{Proposed Caching Policy Design}

From the discussions in Sec. III.C, we understand that the proposed utility maximization can be used in solving many practical and important problems. We thus propose in this section a general solution approach for solving (\ref{eq:optmization_Ut}). Specifically, we propose an approach that iteratively optimizes the caching policies of users. Denote $\mathbf{b}_{k'}=[b_1^{k'},...,b_M^{k'}]^T$ as the policy vector of user $k'$. We iteratively solve the following subproblem for different $k'$ by fixing other users' caching policies:
\begin{subequations}\label{eq:optmization_Ut_LP}
\begin{eqnarray}
\displaystyle{\max_{\mathbf{b}_{k'}}} &U_{\text{LP}}^{k'}=U_{\text{net}}(\mathbf{b}_1,...,\mathbf{b}_{k'},...,\mathbf{b}_K) \label{eq:optmization_Ut_LP_a} \\
\text{subject to} & \sum_{m=1}^M b_m^{k'}=S, \label{eq:optmization_Ut_LP_b}\\
& 0\leq b_m\leq 1,\forall m. \label{eq:optmization_Ut_LP_c}
\end{eqnarray}
\end{subequations}
When $k'\in \mathcal{U}_A$, we obtain
\begin{equation}\label{Ut_reformulation_sum_LP_1}
\begin{aligned}
U_{\text{LP}}^{k'}&=\sum_{k\in\mathcal{U}_A} \frac{w_k U_{\text{D}}}{K_{\text{A}}}+(U_{\text{B}}-U_{\text{D}})\sum_{m=1}^M \sum_{k\in \mathcal{U}_A} \frac{w_ka_{m,k}}{K_{\text{A}}}\left[\prod_{l\in \mathcal{U},l\neq k'}(1-b_m^lL_{k,l})\right]\\
&+(K_{\text{A}}U_{\text{S}}-U_{\text{D}})\left[\sum_{m=1}^M\sum_{k\in \mathcal{U}_A,k\neq k'} \frac{w_ka_m^kb_m^k}{K_{\text{A}}}\right]\\
&-\sum_{m=1}^M b_m^{k'} \left((U_{\text{B}}-U_{\text{D}})\sum_{k\in \mathcal{U}_A} \frac{w_ka_{m,k}}{K_{\text{A}}}L_{k,k'}\left[\prod_{l\in \mathcal{U},l\neq k'}(1-b_m^lL_{k,l})\right]+(U_{\text{D}}-K_{\text{A}}U_{\text{S}})\frac{w_{k'}a_m^{k'}}{K_{\text{A}}}\right);
\end{aligned}
\end{equation}
when $k'\in \mathcal{U}_I$, we obtain
\begin{equation}\label{Ut_reformulation_sum_LP_2}
\begin{aligned}
&U_{\text{LP}}^{k'}=\sum_{k\in\mathcal{U}_A} \frac{w_k U_{\text{D}}}{K_{\text{A}}}+(U_{\text{B}}-U_{\text{D}})\sum_{m=1}^M \sum_{k\in \mathcal{U}_A} \frac{w_ka_{m,k}}{K_{\text{A}}}\left[\prod_{l\in \mathcal{U},l\neq k'}(1-b_m^lL_{k,l})\right]+\\
&(K_{\text{A}}U_{\text{S}}-U_{\text{D}})\left[\sum_{m=1}^M\sum_{k\in \mathcal{U}_A} \frac{w_ka_m^kb_m^k}{K_{\text{A}}}\right]-\sum_{m=1}^M b_m^{k'} \left((U_{\text{B}}-U_{\text{D}})\sum_{k\in \mathcal{U}_A} \frac{w_ka_{m,k}}{K_{\text{A}}}L_{k,k'}\left[\prod_{l\in \mathcal{U},l\neq k'}(1-b_m^lL_{k,l})\right]\right).
\end{aligned}
\end{equation}
Note that (\ref{Ut_reformulation_sum_LP_1}) and (\ref{Ut_reformulation_sum_LP_2}) are simply reformulations of (\ref{Ut_reformulation_sum}), in which we isolate the terms that contain the variables to be optimized. From (\ref{Ut_reformulation_sum_LP_1}) and (\ref{Ut_reformulation_sum_LP_2}), we can see that (\ref{eq:optmization_Ut_LP}) is a linear program.

General linear program solvers could be applied to solve (\ref{eq:optmization_Ut_LP}). However, we provide here a more insightful and efficient approach via the analytical closed-form expressions in (\ref{Ut_reformulation_sum_LP_1}) and (\ref{Ut_reformulation_sum_LP_2}). By letting  
\begin{equation}
U_{\text{LP,S}}^{k',m}=\left\{
\begin{aligned}
& \left((U_{\text{D}}-U_{\text{B}})\sum_{k\in \mathcal{U}_A} \frac{w_ka_{m,k}}{K_{\text{A}}}L_{k,k'}\left[\prod_{l\in \mathcal{U},l\neq k'}(1-b_m^lL_{k,l})\right]+(K_{\text{A}}U_{\text{S}}-U_{\text{D}})\frac{w_{k'}a_m^{k'}}{K_{\text{A}}}\right),&k'\in\mathcal{U}_A,\\
&\left((U_{\text{D}}-U_{\text{B}})\sum_{k\in \mathcal{U}_A} \frac{w_ka_{m,k}}{K_{\text{A}}}L_{k,k'}\left[\prod_{l\in \mathcal{U},l\neq k'}(1-b_m^lL_{k,l})\right]\right),&k'\in\mathcal{U}_I,
\end{aligned}
\right.
\end{equation}
we notice that maximizing $U_{\text{LP}}^{k'}$ is equivalent to maximizing
\begin{equation}\label{eq:Ut_reformulation_sum_LP_2}
\sum_{m=1}^M b_m^{k'} U_{\text{LP,S}}^{k',m}.
\end{equation}
Then, observe that the optimal solution of (\ref{eq:Ut_reformulation_sum_LP_2}), subject to constraints (\ref{eq:optmization_Ut_LP_b}) and (\ref{eq:optmization_Ut_LP_c}), can be obtained by allocating the cache space to the terms offering larger payoffs. Thus, the optimal solution of (\ref{eq:optmization_Ut_LP}) is expressed as
\begin{equation}\label{eq:solution_Ut_LP}
(b_m^{k'})^*=\left\{
\begin{aligned}
&1,&m\in\Phi_{k'},\\
&0,&\text{otherwise},
\end{aligned}
\right.
\end{equation}
where $\Phi_{k'}=\lbrace m:U_{\text{LP,S}}^{k',m} \text{ is among the }S \text{ largest of all }U_{\text{LP,S}}^{k',m}\rbrace$.
By iteratively solving (\ref{eq:optmization_Ut_LP}) via (\ref{eq:solution_Ut_LP}) for different $k'$ until convergence, the caching policy design problem in (\ref{eq:optmization_Ut}) can be effectively solved. Denote $\mathcal{B}_k=\lbrace (b_1^k,...,b_M^k)^T: \sum_{m=1}^M b_m^k=S; 0\leq b_m^k\leq 1,\forall m  \rbrace$. The solution approach is summarized in Alg. \ref{alg:BCD}. Since (\ref{eq:solution_Ut_LP}) suggests that the probability for a user to cache file $m$ is either $1$ or $0$, we actually eliminate the probabilistic interpretation and attain the deterministic policies of users. To characterize the performance of the proposed solution approach, we provide the following theorem:

{\em Theorem 1:} Alg. \ref{alg:BCD} is monotonically non-decreasing at each iteration and can converge to a stationary point if each iteration provides a unique maximizer.\footnote{If the maximizer is not unique, then we will encounter a tie between different $U_{\text{LP,S}}^{k',m}$, which is generally unlikely, as users have different preferences on different files. Thus, such a unique maximizer assumption is mild.}
\begin{proof}
See Appendix B.
\end{proof}

\begin{algorithm}
\caption{Iterative User-Based Caching Policy Design}
\label{alg:BCD}
\begin{algorithmic}
\item{At iteration $r$, choose a user $k'$ and update}
\State{$\mathbf{b}_{k'}^{r+1}=\displaystyle{\arg\max_{\mathbf{b}_{k'}\in\mathcal{B}_{k'}}}U(\mathbf{b}_1^{r},...,\mathbf{b}_{k'-1}^{r},\mathbf{b}_{k'},\mathbf{b}_{k'+1}^{r},...,\mathbf{b}_{K}^{r})$}
\State{$\mathbf{b}_k^{r+1}=\mathbf{b}_{k}^{r},\forall k\neq k'$}
\end{algorithmic}
\end{algorithm}

Finally, we note that although the proposed design needs coordination between users, the users can independently run the proposed Alg. \ref{alg:BCD} since they know the other users' individual preferences. In other words, Alg. \ref{alg:BCD} can actually be implemented in a decentralized manner, given that the users perfectly know the other users' individual preferences.

\subsection{Complexity Analysis of the Proposed Caching Policy Design}
From theorem 1, we can already understand the performance and convergence of the proposed design. Now, we analyze its complexity. Observe that the proposed design is based on the iterative algorithm in Alg. 1. Thus, the complexity comes from the computation at each iteration and the number of iterations required for convergence. At each iteration, the main computational complexity comes from computing for $U_{\text{LP,S}}^{k,m},\forall m$ and sorting $U_{\text{LP,S}}^{k,m},\forall m$. Then, we note that in terms of the total number of additions and multiplications, the complexity order when computing for $U_{\text{LP,S}}^{k,m},\forall m$ is $\mathcal{O}\left(MK^2\right)$; the complexity order of sorting is $\mathcal{O}(M\log M)$. As a result, the overall complexity order at each iteration becomes $\mathcal{O}\left(MK^2+M\log M\right)$. 

Regarding the number of iterations for the convergence, the general analytical expression is intractable; thus, we run simulations to understand how many iterations we would need in practice. For the simulations, we consider the same setup as that shown in Fig. \ref{fg:Fig_3} and evaluate the proposed throughput-based design (see Sec. V for the details of the simulation setup). The convergence of a single user's caching policy does not necessarily imply convergence of all users. Hence, we test the stopping criterion after updating the caching policies of all users in order to guarantee the convergence of all users. This is given as
\begin{equation}
\sum_{m=1}^M\sum_{k=1}^K \vert b_m^k \vert^2 \leq 10^{-4}.
\end{equation} 
Therefore, the number of iterations would be the multiple of $K$, i.e., the number of iterations can only be $K, 2K,3K,...,$ etc. We consider three different numbers of users for the simulations: $K=10,20,30$. The results show that the proposed design can converge within $10K$ number of iterations for more than $99$\% of the cases. In practice, this indicates fast convergence.

\section{Numerical Results}
This section provides simulation results to validate the analysis, evaluate the proposed designs, compare between different designs, and provide insights.

\subsection{Simulation Setup}
We evaluate here the performance of a cluster that covers a square area and has side length $D$. In the simulations, we assume that the users are uniformly distributed within the cluster. Unless otherwise indicated, we assume that the users adopt random-push scheduling. Likewise, we assume that users have equal weights, i.e., $w_k=1,\forall k\in\mathcal{U}_A$. We consider a practical channel model for D2D links, which consists of the pathloss, shadowing, Rayleigh fading, and Gaussian noise. The path-loss model of the D2D link between users $k$ and $l$ is described as \cite{WireCom:Molisch,Ji:Dcache}
\begin{equation}
20\log_{10}\frac{4\pi d_0}{\lambda_c}+10\alpha \log_{10}\frac{d_{k,l}}{d_0},
\end{equation}
where $d_0=10$ m is the breakpoint distance, $\lambda_c=\frac{3\times 10^8}{f_c}$ m, where $f_c=2$ GHz is the carrier frequency, $\alpha=3.68$ is the path-loss exponent, and $d_{k,l}$ is the distance between users $k$ and $l$. The shadowing is modeled by a lognormal distribution with mean $\mu_{\text{dB}}=0$ dB and standard deviation $\sigma_{F}=8$ dB, and the small-scale fading is Rayleigh distributed. We assume that the noise power spectral density is $N_0=-174$ dBm/Hz. We denote $E_{\text{D}}$ as the transmission power of the device and $\text{SNR}_{\text{min}}=5$ dB as the minimum SNR requirement for a successful transmission of a D2D link. Thus, $R_{\text{min}}=\log_2(1+3.16)$ is the minimum transmission rate of a D2D link. We then let $T_{\text{D}}=B_{\text{D}}R_{\text{min}}$ be the throughput of a D2D link, where $B_{\text{D}}=20$ MHz is the bandwidth of a D2D link. We assume that a BS link will always exist whenever a user is scheduled to use it. Since the BS must supply the users in many clusters, we assume that a BS link can share only $\frac{1}{100}$ of the BS resources. Hence, the transmission power of a BS link is $E_{\text{B}}=26$ dBm, which is $\frac{1}{100}$ of the total 46 dBm of the BS power. Similarly, the bandwidth of a BS link is $B_{\text{B}}=200$ kHz, which is $\frac{1}{100}$ of the total $20$ MHz bandwidth. We thus let $T_{\text{B}}=B_{\text{B}}R_{\text{min}}$. Note that a $200$-kHz bandwidth is enough to transmit a low-resolution video, e.g., 360p. We assume that there is no cost when users obtain the desired file from their local caches, and we let $T_{\text{S}}=2T_{\text{D}}$ to indicate the slightly better quality of the video when self-caching is possible.\footnote{Although we can immediately obtain the file when it is in the local cache, the throughput is bounded by the rate at which the user watches the file. Also, mathematically, we should not let $T_{\text{S}}$ go to infinity if we want meaningful results.} For simplicity, we assume  that the energy cost is purely determined by the radio frequency energy required for transmission; access to storage and coding/decoding is assumed to be negligible in comparison. Thus, based on the above setup, we obtain $C_{\text{B}}=E_{\text{B}}$, $C_{\text{D}}=E_{\text{D}}$, and $C_{\text{S}}=0$. Therefore, the EE of the network is $\text{EE}_{\text{net}}=\frac{T_{\text{net}}}{C_{\text{net}}}$ according to the definition in Sec. III.C.5.

We consider $M=1000$ for all simulations. To obtain the individual preferences of users and the corresponding system popularity for the simulations, we use the generator described in \cite{Lee:Indi_pre_model_ToN} to generate individual preference probabilities of $20000$ users to form a dataset. Then, for each realization of the simulations, we randomly select users in the generated dataset for evaluation. As such, the system popularity of the simulations is simply the average of individual preferences of all $20000$ users in the dataset.

In the following, we show the benefits of exploiting the individual preferences by comparing between designs with and without using individual preferences. In other words, we compare between the network performances where the proposed design is implemented either by using the knowledge of individual preference probabilities as in Sec. IV or simply by using the system-wide popularity distribution. In the latter case, the individual preference probabilities of all users in (\ref{eq:optmization_EE}) are replaced by the global (system) popularity distribution, i.e., all users assume the same preference probabilities described by the global popularity distribution. 

\subsection{Effects of Individual Preferences}
In this subsection, we validate the analytical results provided in Sec. III and show the benefits of using information regarding individual preferences. For all simulations in this subsection, we adopt $D=80$ m and $E_{\text{D}}=20$ dBm. In the figures, the results of the proposed design that uses individual preferences is labeled with ``$+$ Individual''; the design that uses global popularity distribution is labeled with ``$+$ Global.'' 

\begin{figure}
\mbox{
\hspace{-45pt}
\begin{subfigure}{0.6\textwidth}
\includegraphics[width=\textwidth]{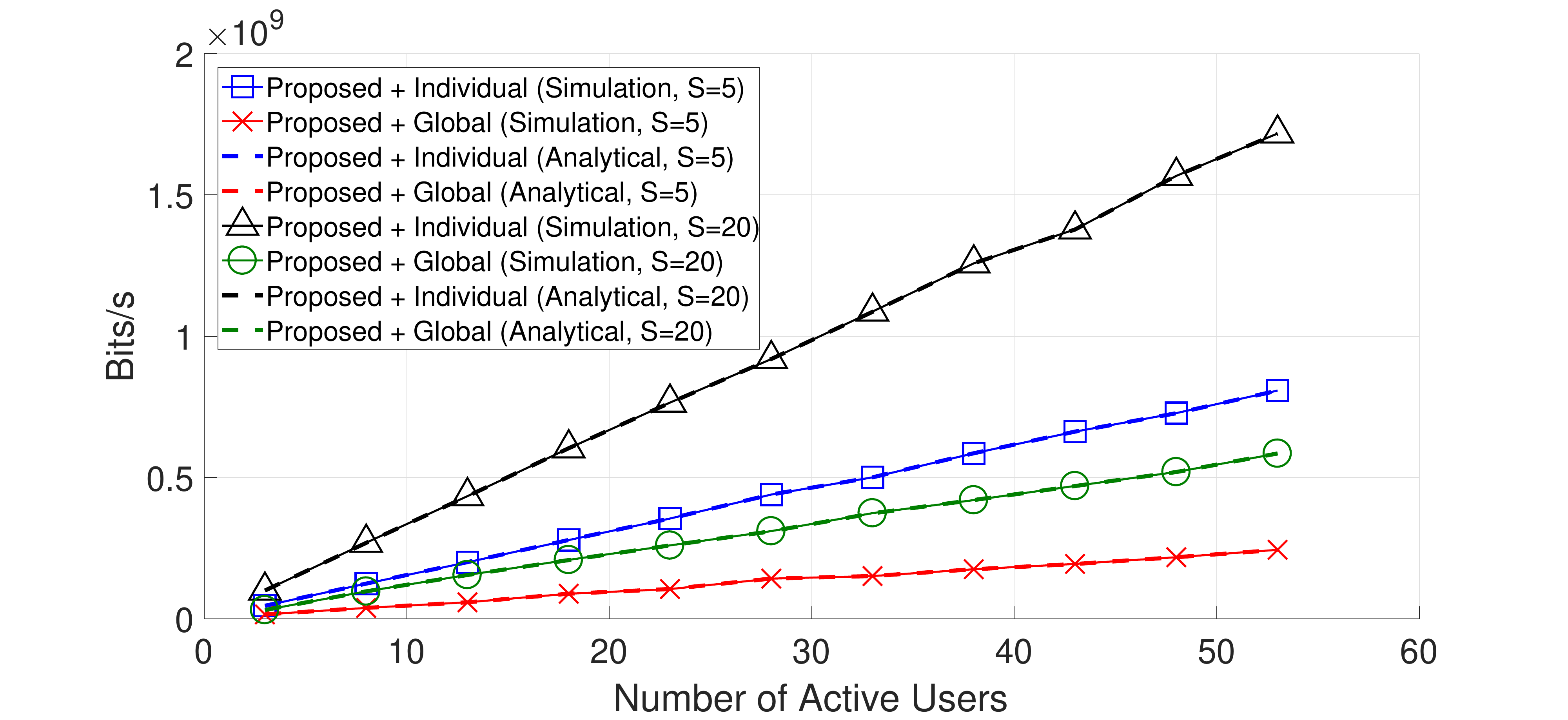}
\caption{Throughput.}
\end{subfigure}
\hspace{-30pt}
\begin{subfigure}{0.6\textwidth}
\includegraphics[width=\textwidth]{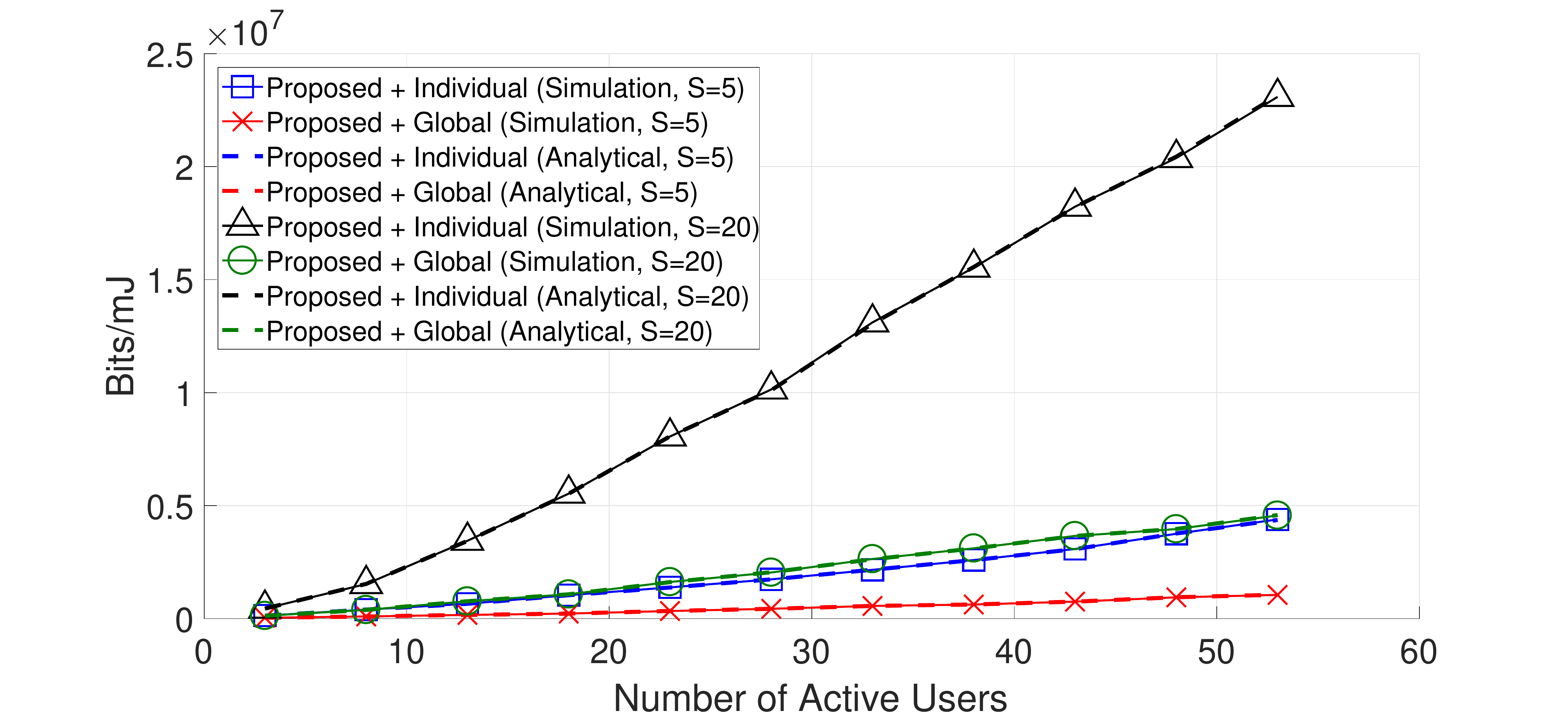}
\caption{EE.}
\end{subfigure}
}
\vspace{-10pt}
\caption{Comparisons between analytical and simulated results in terms of throughput and EE.}
\label{fg:Fig_1}
\vspace{-25pt}
\end{figure}

We first verify our analytical formulations and show the efficacy of using individual preferences. In Fig. \ref{fg:Fig_1}, we consider both $S=5$ and $S=20$ and no inactive users ($K_{\text{I}}=0$), and we evaluate the proposed design in terms of the throughput and EE. When evaluating the throughput in \ref{fg:Fig_1}(a), the proposed design is used for maximizing the throughput; when evaluating the EE in \ref{fg:Fig_1}(b), the proposed design is used for maximizing the EE. The curves labeled with ``Analytical'' are directly computed from expressions in Sec. III; the curves with ``Simulations'' are results of Monte Carlo simulations. We observe that the analytical results match the simulations very well, thereby validating our derivations in Sec. III. Moreover, we see that the design that exploits individual preferences significantly outperforms the corresponding design that does not use individual preferences.

\begin{figure}
\center
\includegraphics[width=0.6\textwidth]{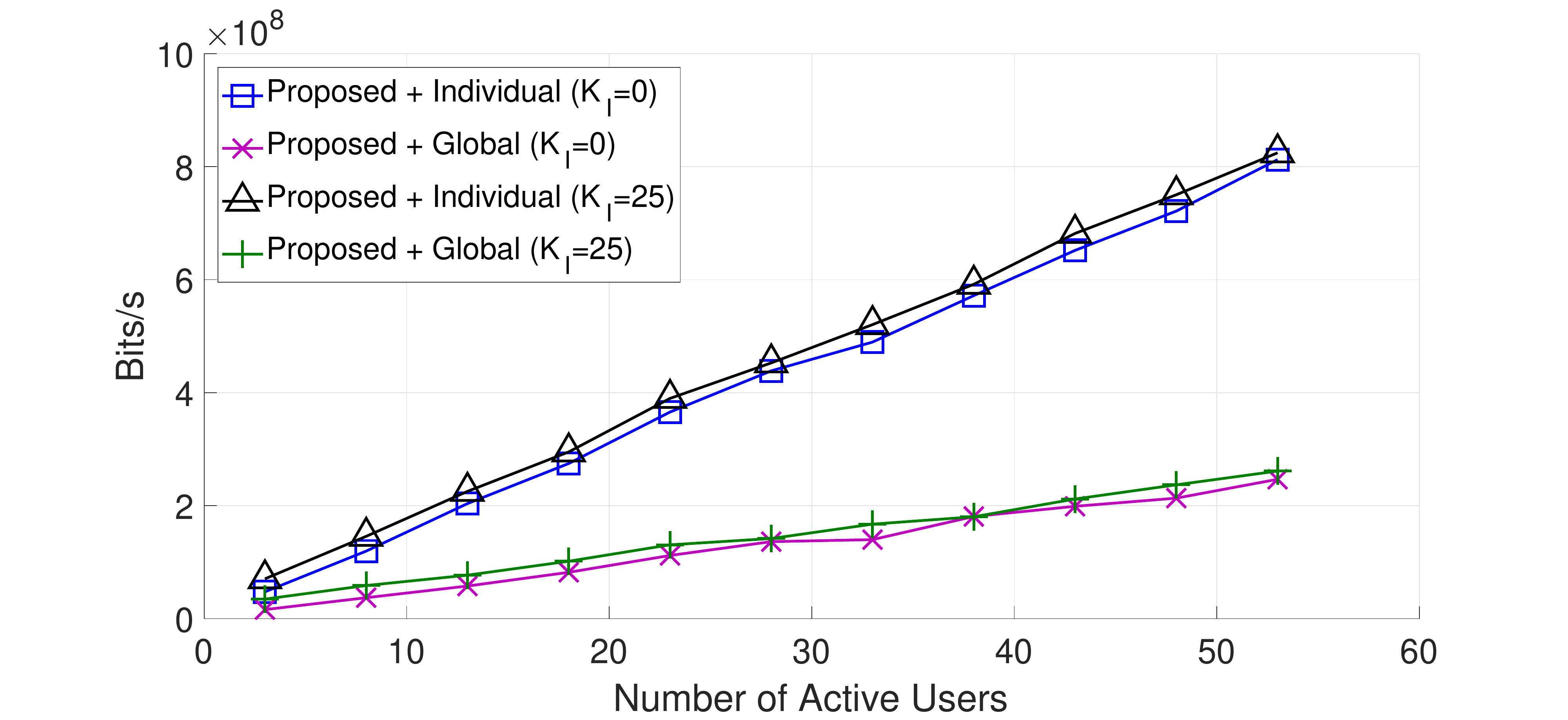}
\caption{Comparisons between networks with different numbers of inactive users in terms of throughput.}
\vspace{-25pt}
\label{fg:Fig_2}
\end{figure}
We next show the impact of inactive users by observing the throughput difference. In Fig. \ref{fg:Fig_2}, we consider $S=5$ and compare the performances of two networks that have different numbers of inactive users, i.e., $K_{\text{I}}=0$ and $K_{\text{I}}=25$. The curves are generated by using the proposed design that aims to maximize throughput. The results show that the benefits of the inactive users are more significant when the number of active users in a cluster is small. At $K_{\text{A}}=3$, the throughput performance improves by $49\%$ when $K_{\text{I}}=25$; at $K_{\text{A}}=53$, it improves by only $1.5\%$ when $K_{\text{I}}=25$. This indicates that although inactive users can help to improve performance, such improvement becomes insignificant when too many users (in the same cluster) share a single D2D band. This implies that when the number of inactive users is large, we might want to have multiple D2D links \cite{Cai2019multiactivation} to benefits more from the inactive users or adjust the number of users in a cluster by reducing the cluster size. However, the use of either approach should be subject to careful tradeoffs between different aspects, such as interference management, power control, reduction of hit-rate, etc.

\subsection{Tradeoff Behaviors between Different Performance Metrics}
In this subsection, we compare different designs and show the tradeoffs between throughput, EE, and hit-rate. Specifically, in all the following figures, we compare between caching policies obtained by using the proposed design framework in pursuit of different goals, i.e., throughput, EE, hit-rate, and the throughput--hit-rate tradeoff, in terms of throughput, EE, and hit-rate. For the throughput--hit-rate tradeoff design, we design the caching policies of users by maximizing $T_{\text{net}}+\zeta T_{\text{D}} K_{\text{A}}H_{\text{net}}$, i.e., by using $U_{\text{B}}=T_{\text{B}}$, $U_{\text{D}}=T_{\text{D}}+\zeta K_{\text{A}}T_{\text{D}}$, and $U_{\text{S}}=T_{\text{S}}+\zeta T_{\text{D}}$. Such a tradoff design is interpreted as a weighted sum of throughput and hit-rate in which the throughput is rendered the weight $1$ and the hit-rate rendered the weight $\zeta K_{\text{A}}T_{\text{D}}$. Note that the term $T_{\text{D}}$ in the weight of the hit-rate is basically to calibrate between different units. This tradeoff design is then labeled with ``TH-HIT Tradeoff - $\zeta$'' in the figures, where $\zeta$ might be different to indicate different tradeoff behaviors.

\begin{figure}
\centering
\begin{subfigure}{0.6\textwidth}
\includegraphics[width=\textwidth]{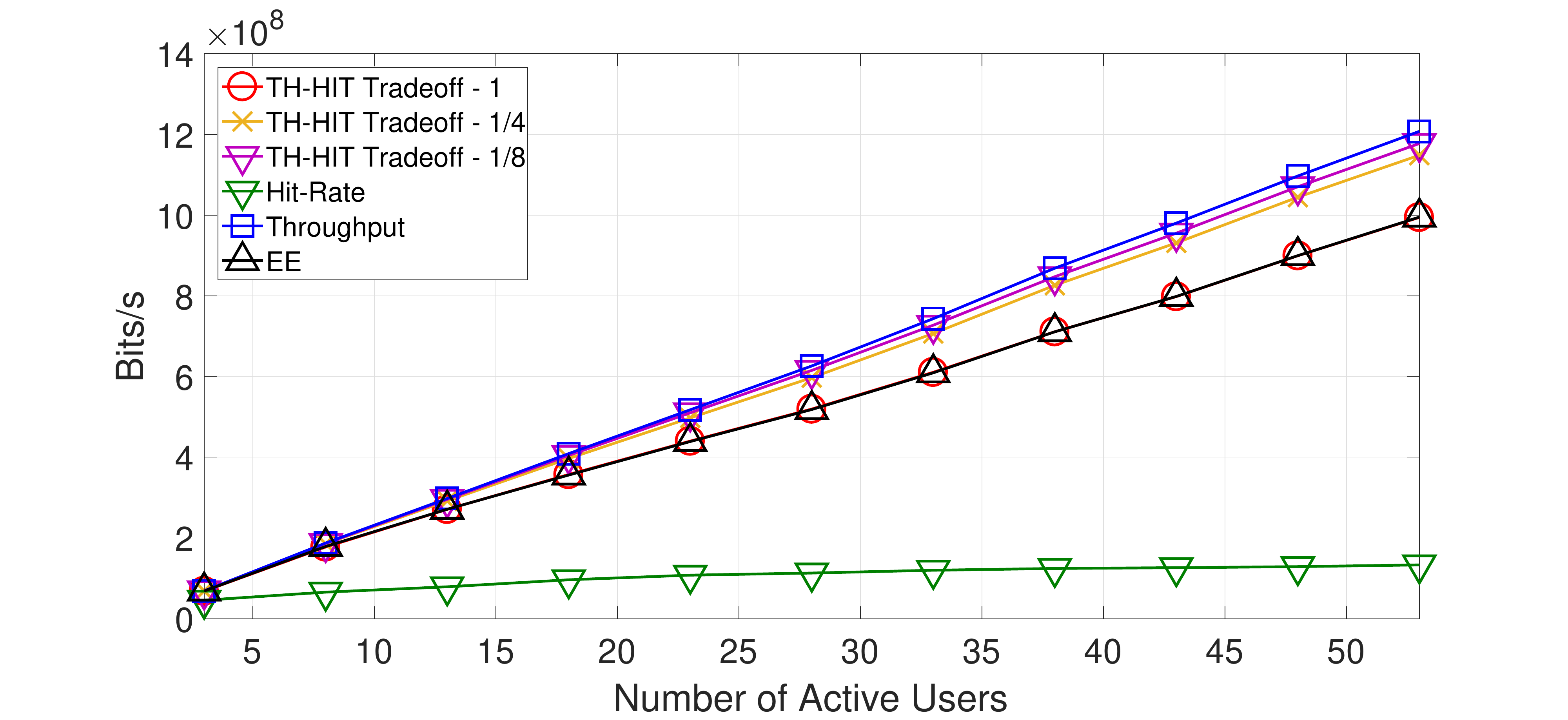}
\caption{Throughput.}
\end{subfigure}\\
\mbox{
\hspace{-45pt}
\begin{subfigure}{0.6\textwidth}
\includegraphics[width=\textwidth]{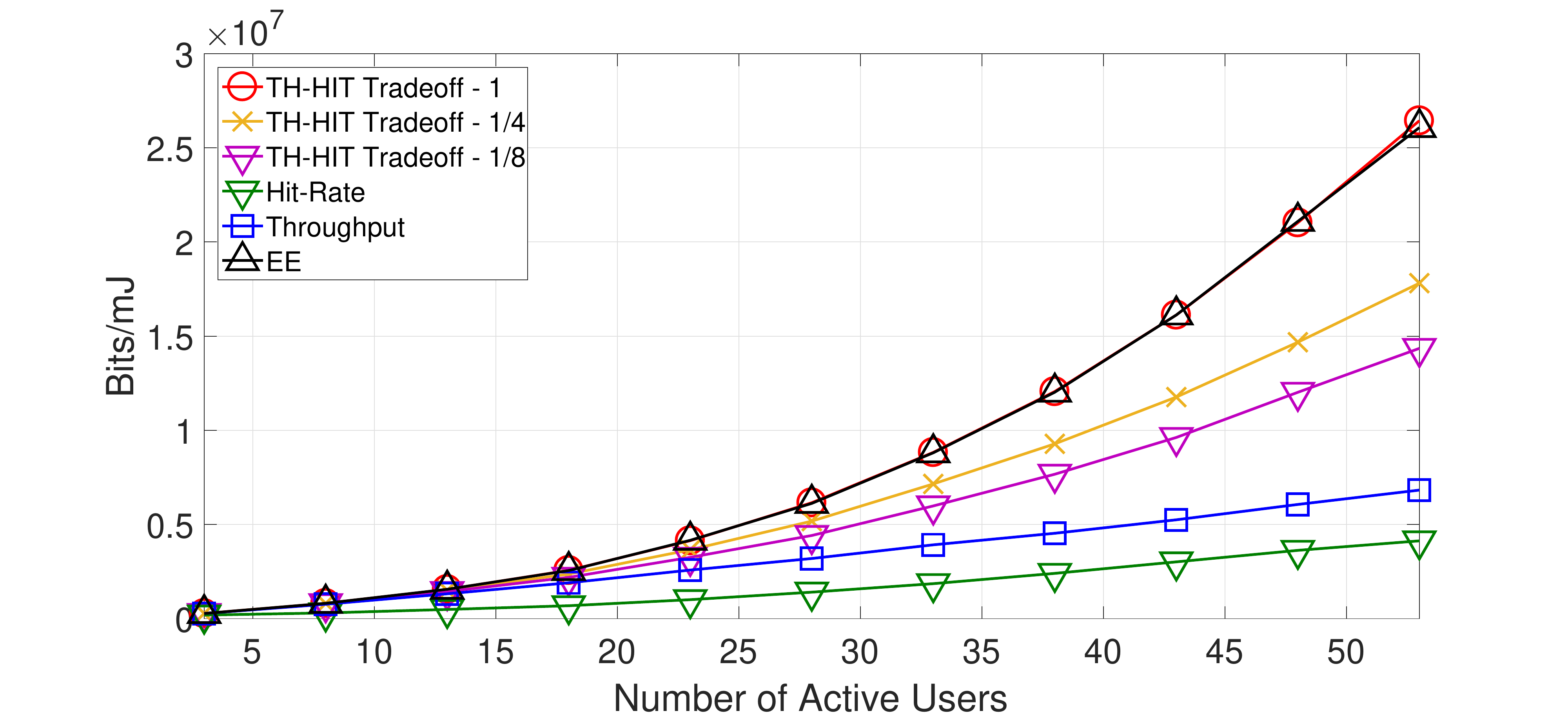}
\caption{EE.}
\end{subfigure}
\hspace{-30pt}
\begin{subfigure}{0.6\textwidth}
\includegraphics[width=\textwidth]{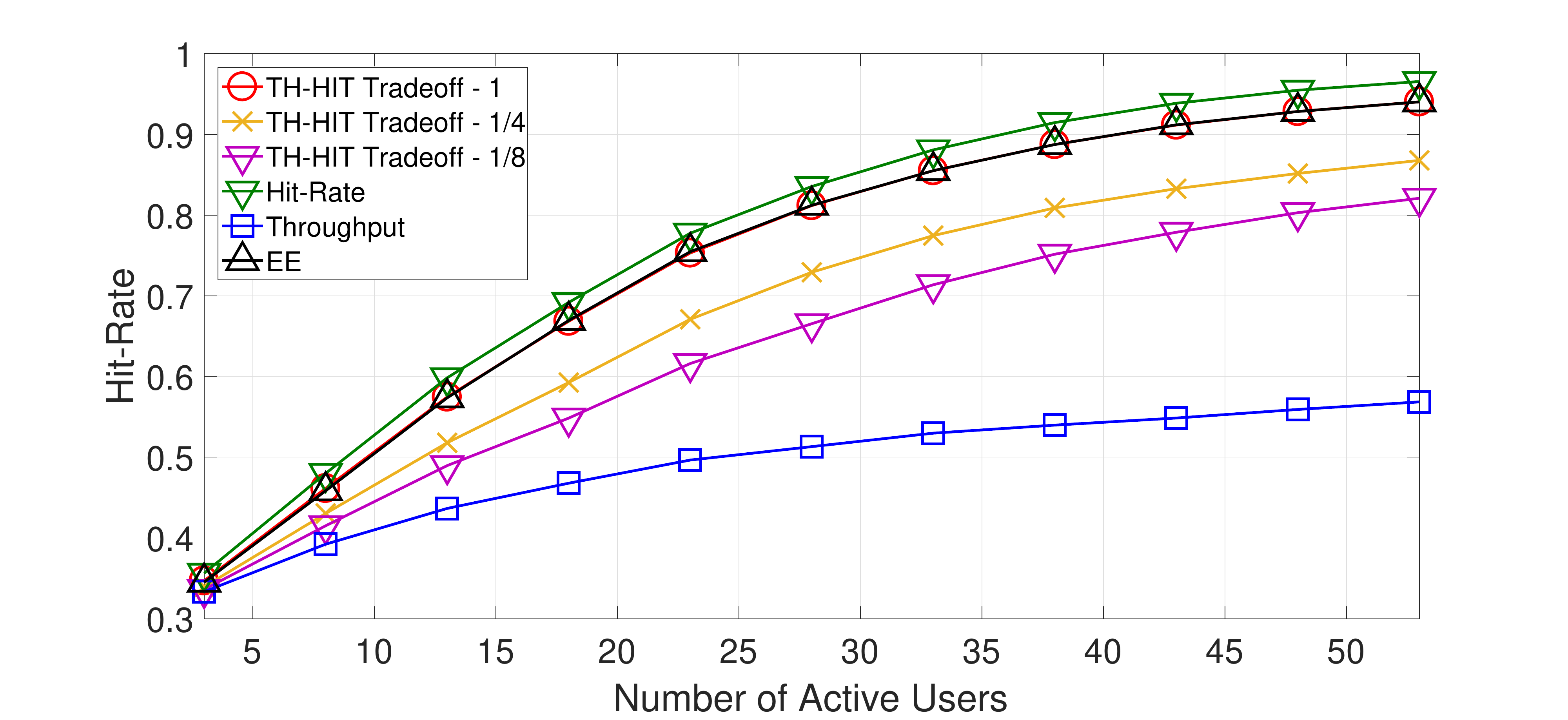}
\caption{Hit-Rate.}
\end{subfigure}
}
\vspace{-10pt}
\caption{Comparisons between different designs in terms of throughput, EE, and hit-rate.}
\label{fg:Fig_3}
\vspace{-15pt}
\end{figure}

Considering $S=10$, $R=80$, $E_{\text{D}}=13$ dBm, and $K_{\text{I}}=0$, we compare different designs in Fig. \ref{fg:Fig_3}. Unsurprisingly, the throughput-based, EE-based, and hit-rate-based designs provide the best throughput, EE, and hit-rate, respectively. The hit-rate-based design provides poor throughput because it does not consider the self-caching gains possibly brought by letting users cache their desired files. In contrast, the throughput-based design is not effective in terms of hit-rate because the design overemphasizes self-caching gains. By striking a balanced viewpoint between throughput and hit-rate, the appropriate throughput--hit-rate tradeoff designs can efficiently trade throughput for hit-rate. This can then significantly improve the hit-rate without degrading the throughput much. Note that by adjusting $\zeta$, we can effectively adjust the tradeoff behavior. Finally, we observe that an energy-efficient caching policy can be obtained by balancing the throughput and hit-rate. In fact, when $\zeta=1$, the throughput--hit-rate tradeoff design performs almost as well as our proposed EE-based design. Also, it is worthwhile to note that when compared to the use of $E_{\text{D}}=20$ dBm, as in Fig. \ref{fg:Fig_1}, the adoption of $E_{\text{D}}=13$ dBm here indeed reduces the power consumption significantly, resulting in much better EE. However, such a transmission power reduction only slightly increases the channel outage.\footnote{The channel outage rate increases by only $0.012$.} This implies the usefulness of a good power control policy of the network.

\begin{figure}
\centering
\begin{subfigure}{0.6\textwidth}
\includegraphics[width=\textwidth]{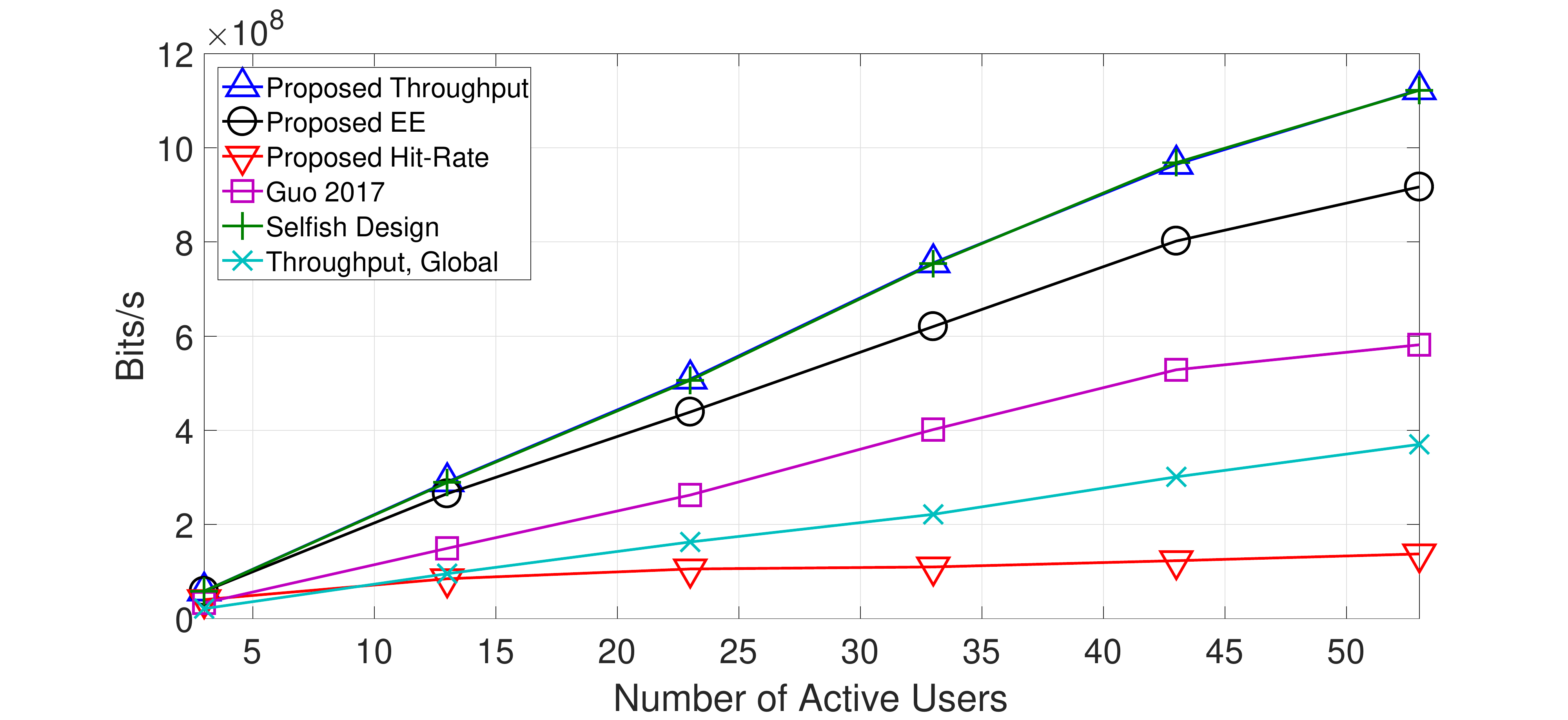}
\caption{Throughput.}
\end{subfigure}\\
\mbox{
\hspace{-45pt}
\begin{subfigure}{0.6\textwidth}
\includegraphics[width=\textwidth]{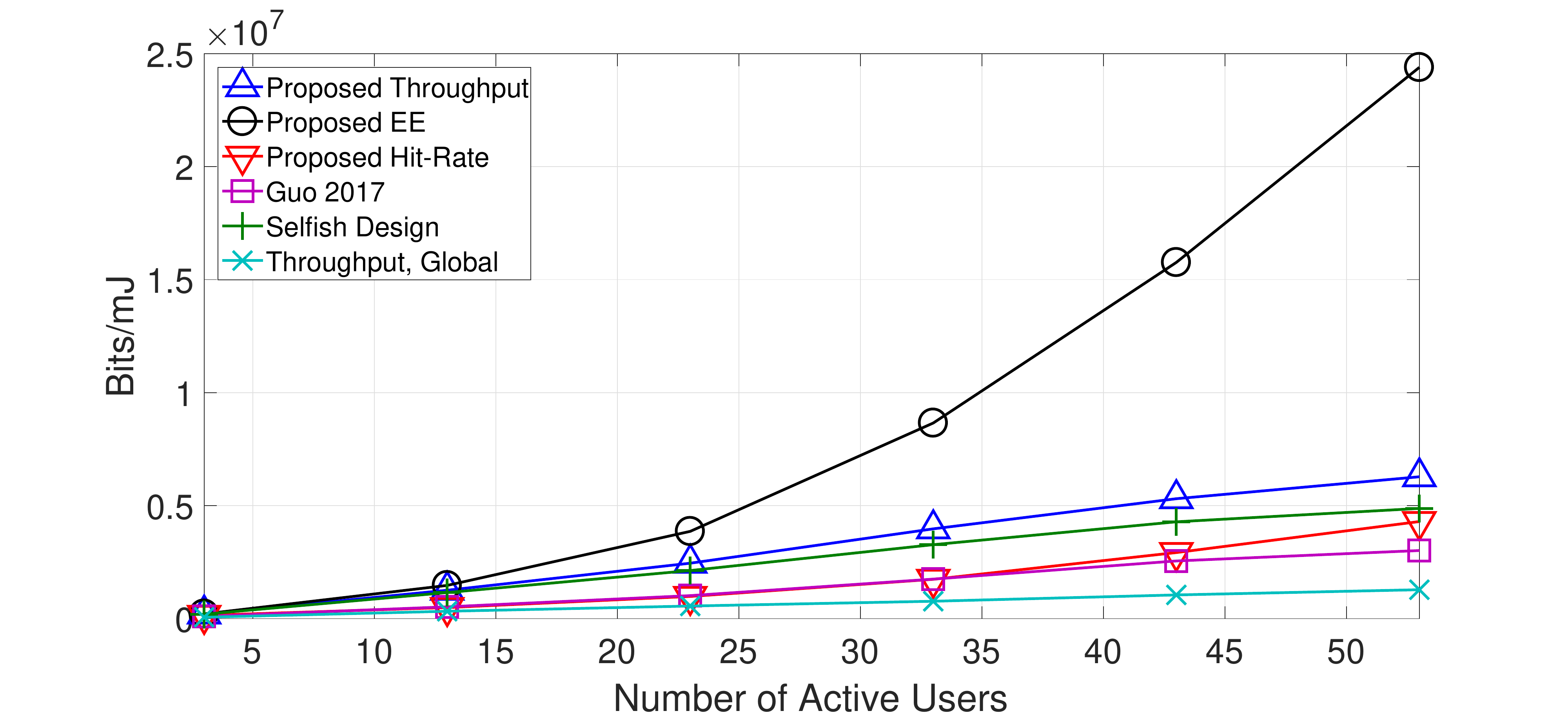}
\caption{EE.}
\end{subfigure}
\hspace{-30pt}
\begin{subfigure}{0.6\textwidth}
\includegraphics[width=\textwidth]{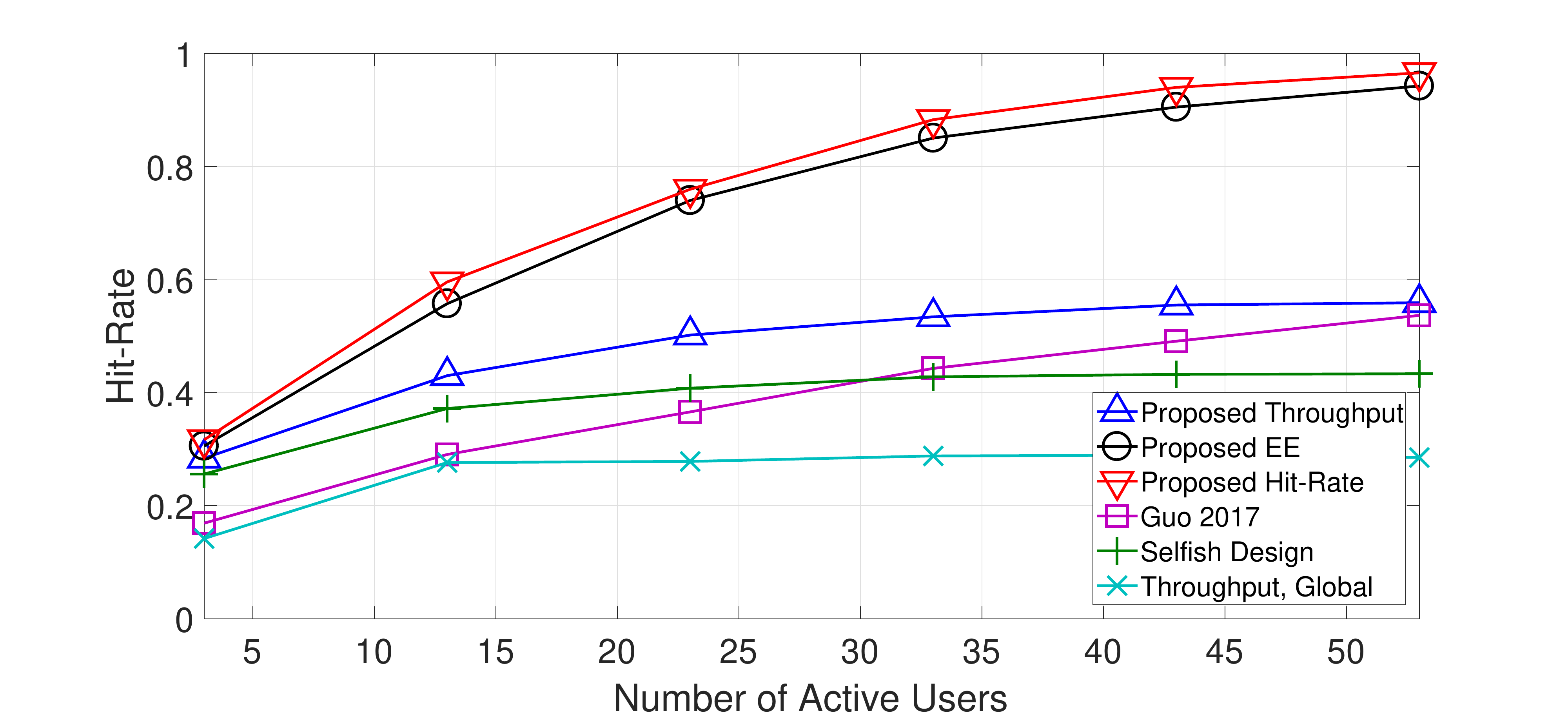}
\caption{Hit-Rate.}
\end{subfigure}
}
\vspace{-10pt}
\caption{Comparisons between different designs in terms of throughput, EE, and hit-rate.}
\label{fg:Fig_4}
\vspace{-15pt}
\end{figure}

We now consider the same setup as featured in Fig. \ref{fg:Fig_3} and compare the proposed design to some other reference designs in Fig. \ref{fg:Fig_4}. Specifically, we compare the proposed design to the baseline selfish design, in which each user selfishly caches the files according to its own preference without considering other users. The selfish design can be considered as an extreme as opposed to the maximum hit-rate design, which maximizes cooperation between users. We also compare the proposed designs to the design that adopts the global popularity distribution, similar to those in Figs. \ref{fg:Fig_1} and \ref{fg:Fig_2}. Furthermore, we compare the proposed designs to Alg. 1 in \cite{Guo:PreCache_2}, which is labeled as ``Guo 2017.'' To adapt the design in \cite{Guo:PreCache_2} to our network model, we make some revisions to it. First, we let each group defined in \cite{Guo:PreCache_2} stand for only a single user. Additionally, we let the cooperation range defined in \cite{Guo:PreCache_2} be the same as the cluster size defined in our paper. Since the design proposed in \cite{Guo:PreCache_2} assumes that a user can make caching decisions only for a single cache space, we implement a naive extension for it by treating each caching space of a user independently and repeatedly using the same policy designed in \cite{Guo:PreCache_2} for every caching space of the user. 

The results show that our proposed designs outperform all the other designs in terms of throughput, EE, and hit-rate. Moreover, the selfish design performs well in terms of throughput, while it performs very poorly in terms of EE and hit-rate. This is because the network throughput can be effectively enhanced by having large local gains when all users are active. However, this near-optimal selfish design cannot hold if there are inactive users. We will see this later in another figure. On the other hand, the selfish design inherently provides very poor hit-rate, leading to poor EE, since the BS links are frequently used.

\subsection{Performance Evaluations with Respect to Cluster Size}

\begin{figure}
\centering
\begin{subfigure}{0.6\textwidth}
\includegraphics[width=\textwidth]{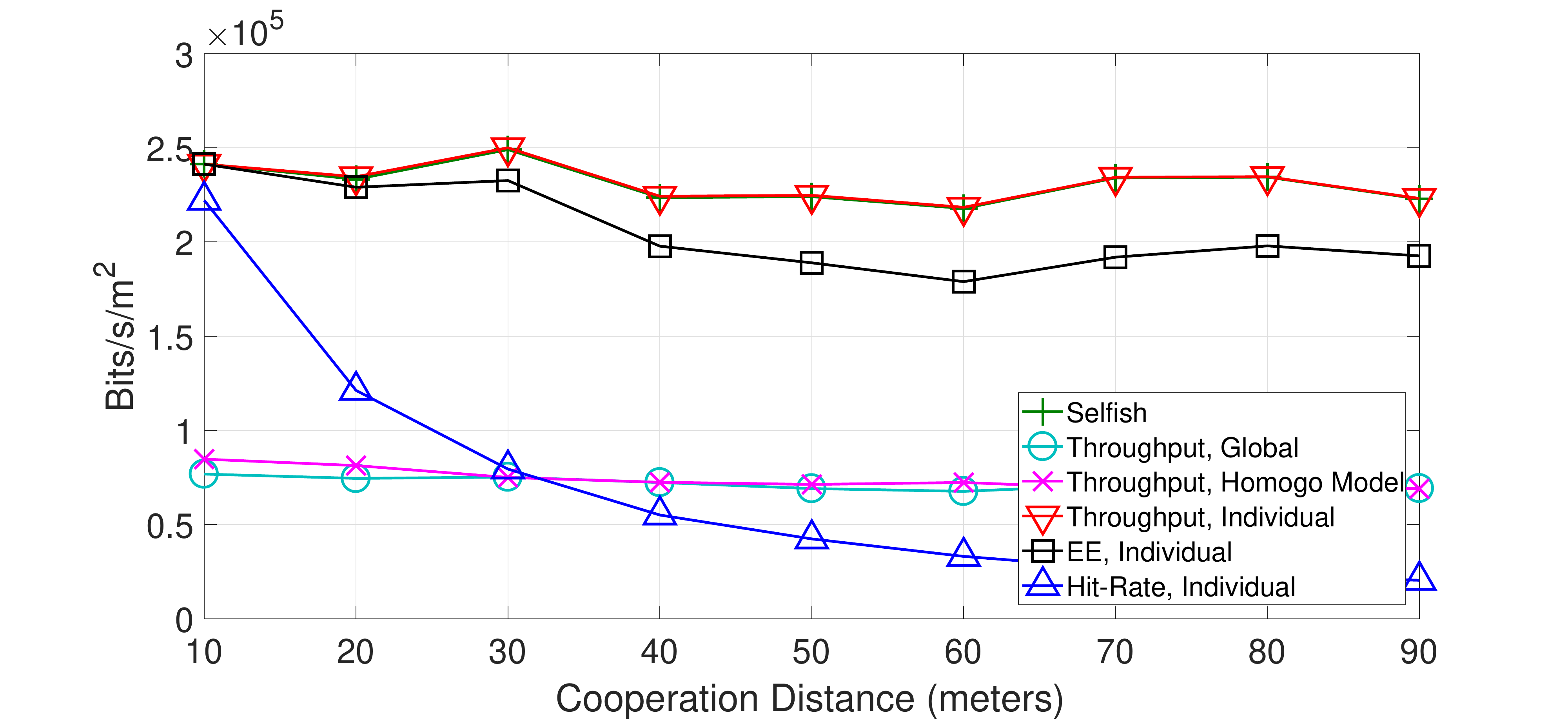}
\caption{Throughput.}
\end{subfigure}\\
\mbox{
\hspace{-45pt}
\begin{subfigure}{0.6\textwidth}
\includegraphics[width=\textwidth]{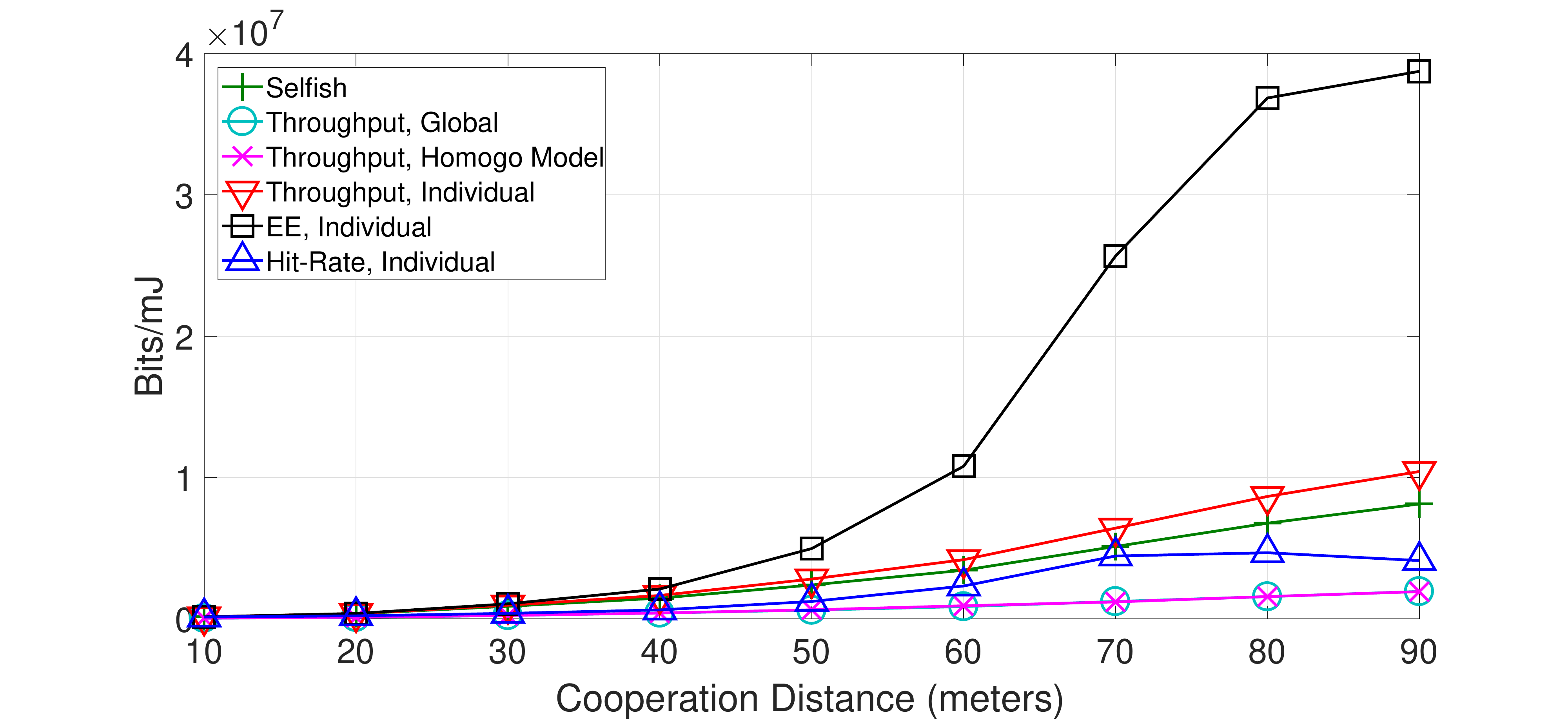}
\caption{EE.}
\end{subfigure}
\hspace{-30pt}
\begin{subfigure}{0.6\textwidth}
\includegraphics[width=\textwidth]{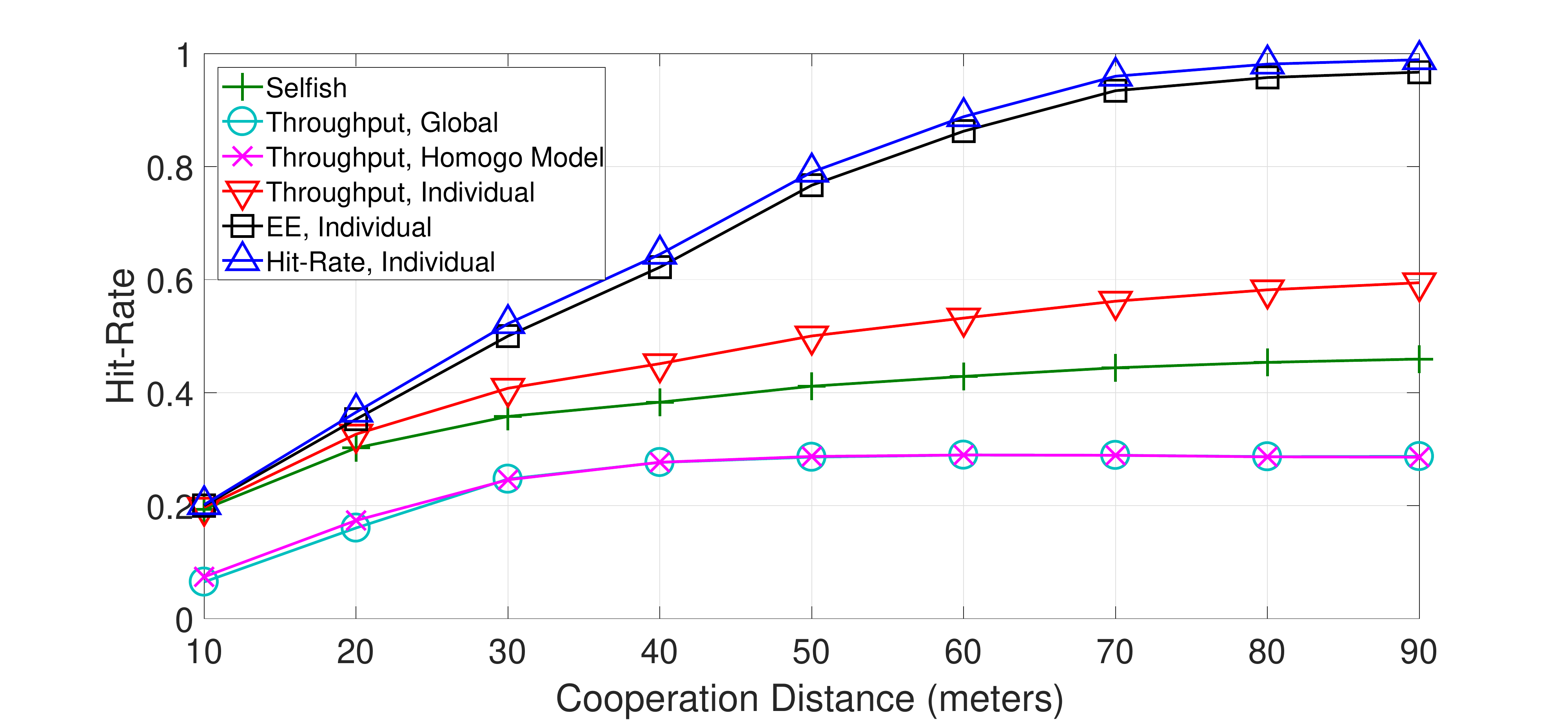}
\caption{Hit-Rate.}
\end{subfigure}
}
\vspace{-10pt}
\caption{Comparisons between different designs in terms of throughput, EE, and hit-rate with respect to cluster size with $\lambda_{\text{A}}=0.01$ and $\lambda_{\text{I}}=0$.}
\label{fg:Fig_5}
\vspace{-30pt}
\end{figure}

\begin{figure}
\centering
\begin{subfigure}{0.6\textwidth}
\includegraphics[width=\textwidth]{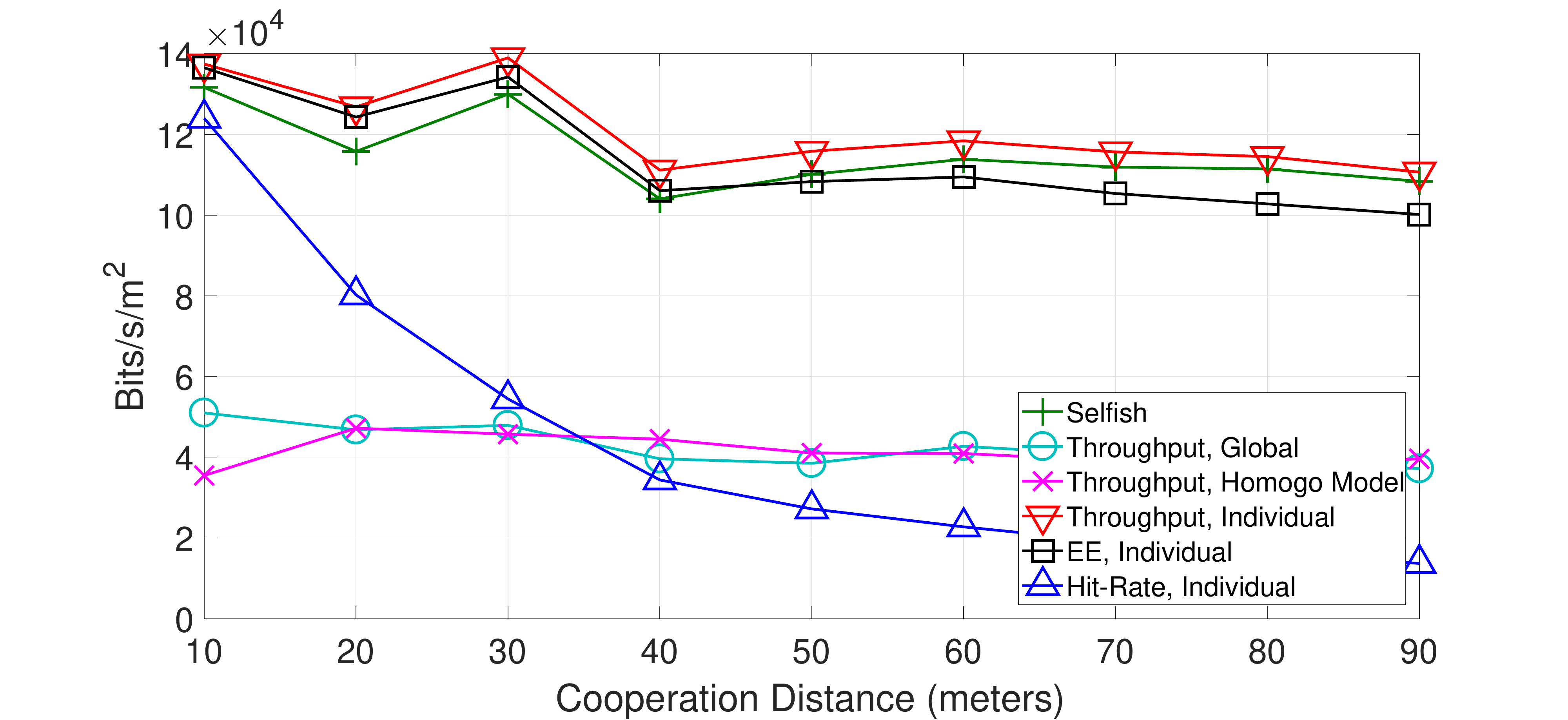}
\caption{Throughput.}
\end{subfigure}\\
\mbox{
\hspace{-45pt}
\begin{subfigure}{0.6\textwidth}
\includegraphics[width=\textwidth]{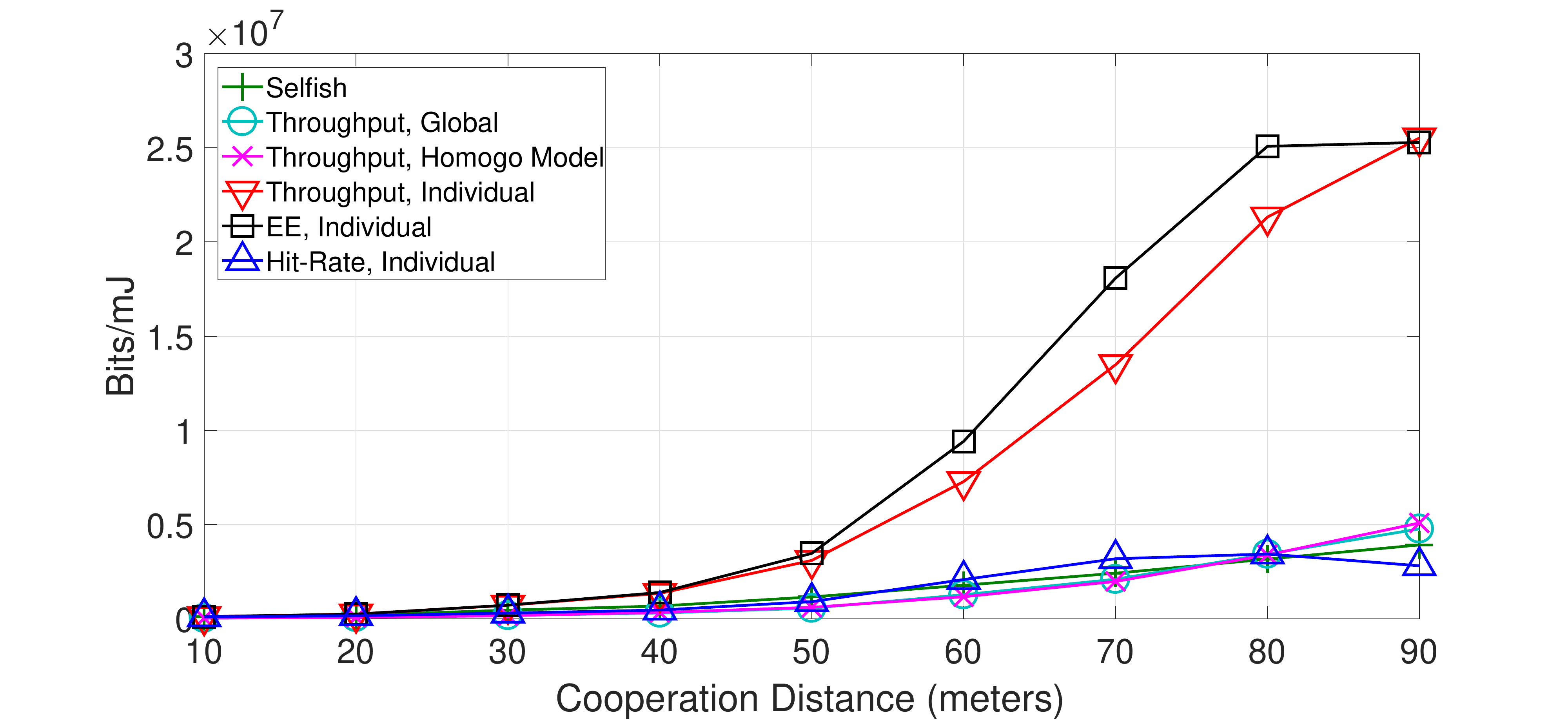}
\caption{EE.}
\end{subfigure}
\hspace{-30pt}
\begin{subfigure}{0.6\textwidth}
\includegraphics[width=\textwidth]{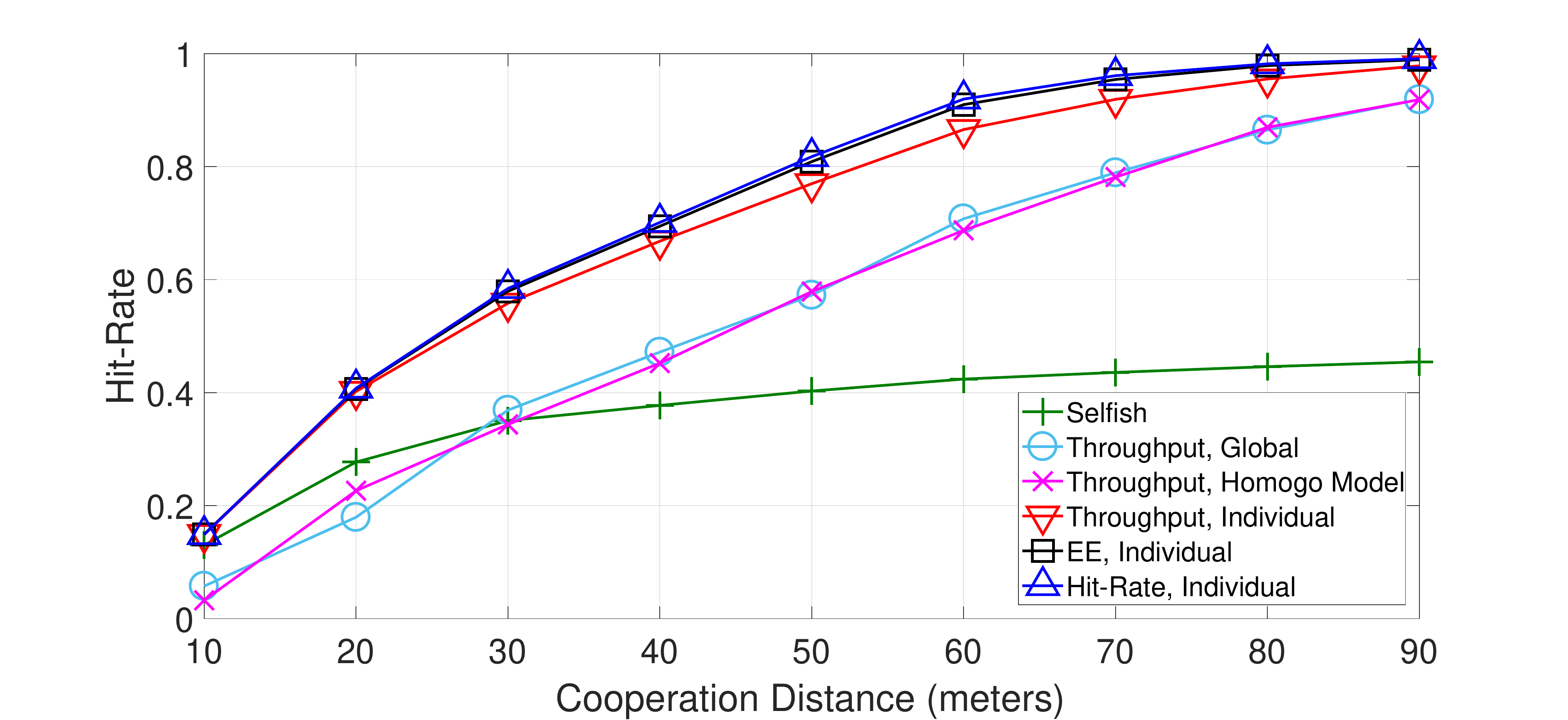}
\caption{Hit-Rate.}
\end{subfigure}
}
\vspace{-10pt}
\caption{Comparisons between different designs in terms of throughput, EE, and hit-rate with respect to cluster size with $\lambda_{\text{A}}=0.005$ and $\lambda_{\text{I}}=0.005$.}
\label{fg:Fig_6}
\vspace{-15pt}
\end{figure}

Fig. \ref{fg:Fig_5} and Fig. \ref{fg:Fig_6} evaluate the proposed designs with respect to the cluster size $D$. A change in the cluster size should be accompanied by a suitable transmission power control of D2D links. Hence, we adopt the power control policy proposed in \cite{Lee:caching} to appropriately manage the average SNR of the received signal and the interference between clusters. This power control policy is:
\begin{equation}\label{eq:pwControl}
E_{\text{D}}=\left[(\sqrt{K}-1)\frac{d}{d_0}\right]^{\alpha}\cdot (\frac{4\pi d_0}{\lambda_{\text{c}}})^2\cdot \nu,
\end{equation}
where $K=16$ is the reuse factor and $\nu=2^{\frac{\alpha}{2}}N_0B_{\text{D}}$ is the maximum allowable interference between clusters.\footnote{The value of $\nu$ is at the level of noise power; hence, for brevity, we ignore the inter-cluster interference in the simulations. Accordingly, we use $\nu$ here to compute only for $E_{\text{D}}$.} Such a power control policy can adjust the transmission power of devices such that the average SNR of the received signal and the inter-cluster interference are almost invariant when changing the cluster size. Since D2D links are expected to exist only for short-distance transmissions, we consider $D\leq 90$ m. This results in $E_{\text{D}}\leq 20$ dBm when we use (\ref{eq:pwControl}) to adjust the power. We consider the Poisson point processes to model the number of active and inactive users in the cluster, where $\lambda_{\text{A}}$ and $\lambda_{\text{I}}$ represent the densities of active and inactive users, respectively. Thus, the numbers of active and inactive users are the random variables described by the Poisson distributions with parameter $\lambda_{\text{A}} D^2$ and $\lambda_{\text{I}} D^2$, respectively. Also, we need to accommodate the fact that a cluster has different numbers of users when $D$ is different. Hence, instead of directly looking at the throughput, we evaluate using the throughput per area ($\text{Bits}/\text{s}/\text{m}^2$). 

Similar to Fig. \ref{fg:Fig_3}, we compare between different caching policies developed through the proposed design, with the aim of maximizing different performance metrics. In addition, we compare with two reference curves in Fig. \ref{fg:Fig_4}. Furthermore, we include an additional reference curve, which adopts the coordinated design with homogeneous modeling (labeled ``Homogo Model'' in Figs. \ref{fg:Fig_5} and \ref{fg:Fig_6}). This curve considers the situation that the policy is designed using global popularity distribution while the users indeed have the same preference, following the global popularity distribution. It thus represents the performance of systems that design and evaluate using the homogeneous modeling employed in previous papers -- we want to see the influences on the performance evaluation of cache-aided D2D networks when changing from a homogeneous modeling to a more practical heterogeneous modeling.

In Fig. \ref{fg:Fig_5}, we consider $S=10$, $\lambda_{\text{A}}=0.01$, and $\lambda_{\text{I}}=0$, i.e., no inactive users. Since the area throughput and EE are influenced by multiple factors, they are not convex/concave functions in general. As such, we see that the area throughput of the throughput-based design fluctuates when $D$ is small; it becomes somewhat flat when $D$ is large. This is because the contribution of the D2D transmission becomes minor as too many users share the same D2D band in a cluster. We also see that the selfish design is relatively effective again because all users are active. As expected, the hit-rate-based design provides the best hit-rate. Meanwhile, the area throughput of the hit-rate-based design continuously decreases with respect to $D$, since it strives to improve the hit-rate without considering the influence of the self-caching gain. In contrast, the throughput-based design again provides the best area throughput, but it is not effective for hit-rate. In terms of EE, the EE-based design outperforms others significantly. Moreover, the optimal point of the EE is at a large cooperation distance because it is necessary to have high hit-rate in order to have large EE; otherwise, the BS needs to serve the users by using more BS links, leading to the smaller throughput and larger power consumption, and thus poor EE.

Through observation, we see that the tradeoff between the area throughput and EE can be attained not only through different caching policies but also through different cluster sizes. Thus, a network designer should consider both the caching policy and cluster size when designing the network. Finally, we argue that exploiting individual preferences is expectedly beneficial. We can see that our proposed system, which considers individual preferences in the design, performs better than the system that operates under the assumption that users have the same preferences, i.e., the curve with the ``Homogo Model'' label. Such a result implies that, rather than being detrimental, the diverse preferences of users for files can actually be used to further improve the network.

In Fig. \ref{fg:Fig_6}, we conduct a similar evaluation as in Fig. \ref{fg:Fig_5}. Here, we adopt $\lambda_{\text{A}}=0.005$ and $\lambda_{\text{I}}=0.005$, i.e., there are some inactive users. We can see that most of the phenomena observed in Fig. \ref{fg:Fig_5} can be observed again here. Since we now have inactive users, they should be cooperative such that we can obtain the optimal throughput while the active users are still fairly selfish. This then distinguishes the selfish design from our proposed design and causes the throughput-based design to perform well in terms of EE and hit-rate. We can actually observe that the difference between the optimal values of the throughput- and EE-based designs in terms of EE is smaller as compared to that in Fig. \ref{fg:Fig_5}. However, the tradeoff between throughput and EE is still significant as we change the cluster size. Finally, we see that the proposed system performs better than the system with pure homogeneous modeling. This again validates our point that having users with diverse preferences is beneficial.

\subsection{Performance Evaluations with Different Schedulers}
Finally, we evaluate the proposed design in clustering networks with two different schedulers. Through this, we can show how the proposed design can help developers design a caching policy for a network that has a very complicated scheduler. Specifically, in addition to evaluating using the random-push scheduler, we evaluate (under the same caching policy) the ``priority-push scheduler'' \cite{Lee:caching}, which functions as follows: all users initially check whether the files in their local caches can satisfy their requests. If yes, then the requests are satisfied; otherwise, they send their requests to the BS. The BS then checks whether there are users that can be satisfied by using D2D links. If yes, then the BS randomly selects one user to be served by the D2D link; otherwise, the BS randomly selects one user from those sending the requests and then serves the user via a BS link. Such a scheduler maximizes the usage of D2D communications. Hence, we can expect that the priority-push network will have higher throughput and better EE than the random-push network. On the other hand, it might be unfair to those users whose preferences are not similar to the mainstream; they might be less likely to be selected and accordingly be served. More importantly, such a complicated scheduler results in an intractable expression for designing caching policies. We demonstrate how to exploit the proposed designs in this work along with some numerical results to guide the designer in obtaining effective designs for it.

\begin{figure}
\centering
\mbox{
\hspace{-45pt}
\begin{subfigure}{0.6\textwidth}
\includegraphics[width=\textwidth]{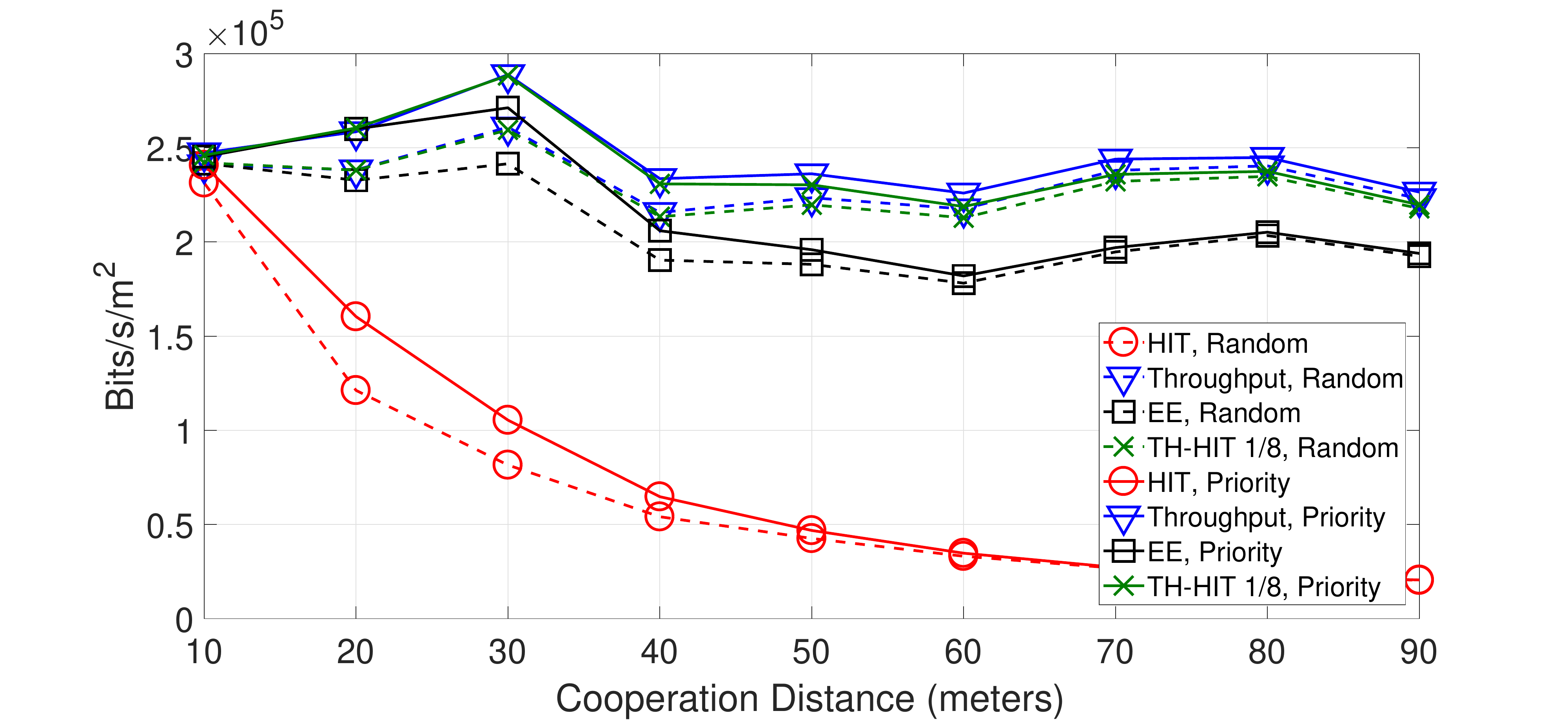}
\caption{Throughput.}
\end{subfigure}
\hspace{-30pt}
\begin{subfigure}{0.6\textwidth}
\includegraphics[width=\textwidth]{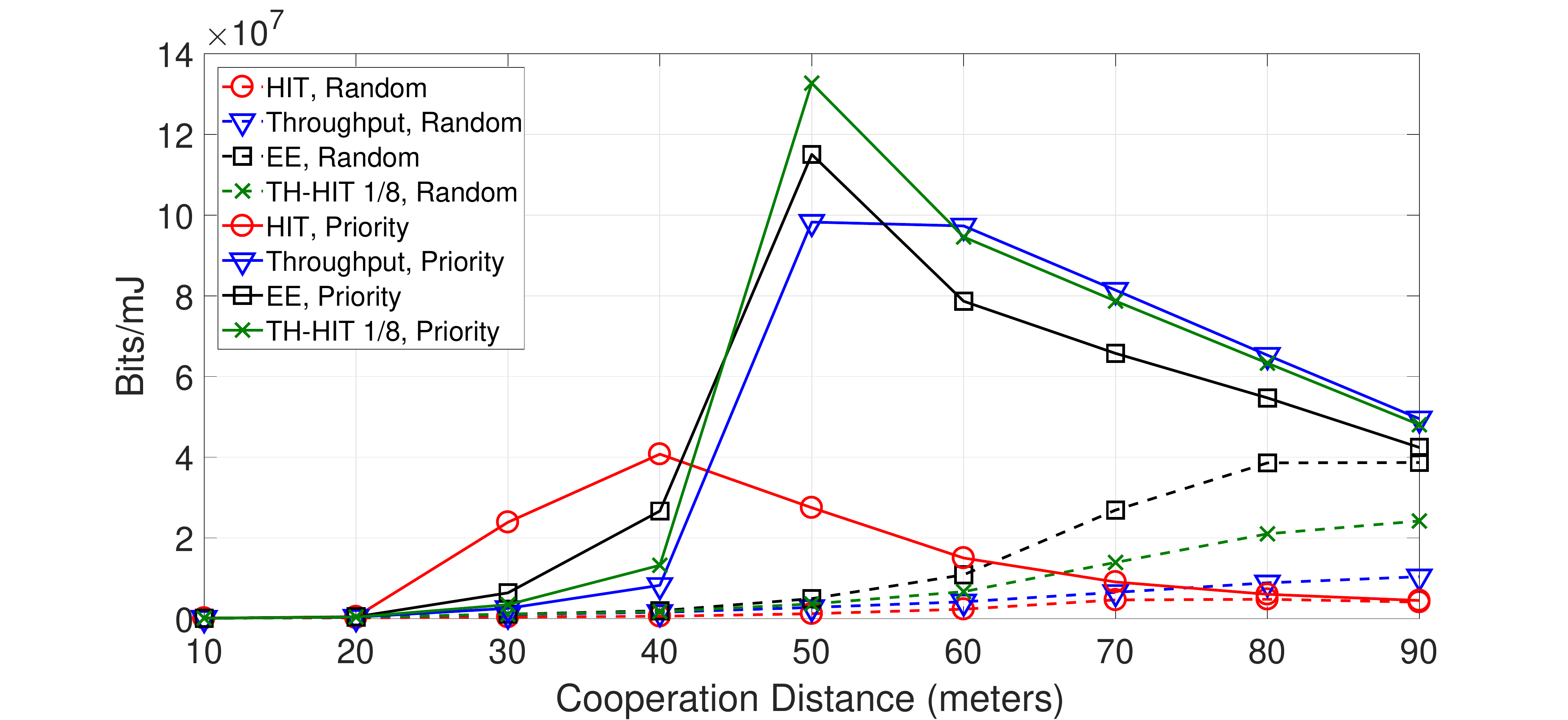}
\caption{EE.}
\end{subfigure}
}
\vspace{-10pt}
\caption{Comparisons between different designs in terms of throughput and EE with respect to cluster size with $\lambda_{\text{A}}=0.01$ and $\lambda_{\text{I}}=0$.}
\label{fg:Fig_7}
\vspace{-20pt}
\end{figure}

In Fig. \ref{fg:Fig_7}, we consider the same setup as in Fig. \ref{fg:Fig_5} and evaluate the proposed design in both networks using the random-push and priority-push schedulers, labeled as ``Random'' (dashed line) and ``Priority'' (solid line), respectively. We observe that the priority-push network generally outperforms the random-push network in terms of the area throughput and EE.\footnote{Since the hit-rate considering the priority-push scheduling is the same as the hit-rate considering the random-push scheduling, we omit the demonstration of the hit-rate for brevity.} Additionally, we observe that in terms of the area throughput, the results of the random-push network can be fairly representative. The results for EE show more subtle effects. We see that the optimal cluster size for the priority-push network is much smaller, which implies that it is unnecessary in the priority-push network to have high hit-rate to obtain the best EE. This is because the priority-push scheduler would schedule a D2D link as long as there exists one, implying that it would have a higher rate for scheduling D2D links than simply the hit-rate -- the probability for at least one user to find the desired file in the D2D network is higher than that of a particular user to find its desired file. Thus, to obtain an energy-efficient design in the priority-push network, we need to choose a design that has smaller cluster size and has lower hit-rate than those that provide the optimal EE in the random-push network. Overall, based on the above results, we conclude that we need to reduce the cluster size and consider various tradeoff designs proposed in this paper in order to obtain an effective design in the priority-push network. Since our proposed tradeoff designs can efficiently evaluate the throughput and hit-rate, such a trial-and-error procedure might not be challenging.

\vspace{-10pt}

\subsection{Summary of the Insights}
{
Here we summarize the insights from our simulation results:
\begin{itemize}
\item It is necessary to consider the influence of users' individual preferences on system design because evaluations done under the assumption that users have the same preferences are not representative of evaluations done while considering individual preferences. Therefore, by considering the effects of individual preferences, the proposed designs can significantly improve network performance.
\item A system optimized for throughput can have significant loss in EE and vice versa. Similarly, a system optimized for hit-rate can have significant loss in throughput and EE and vice versa. However, allowing slight sub-optimality in one performance can significantly improve another performance.
\item Instead of directly optimizing EE, which can be complicated, one alternative approach is to use a throughput--hit-rate tradeoff design, i.e., by balancing between throughput and hit-rate, a design with high EE can be obtained.
\item While directly finding the effective caching policy for the network using priority-push scheduling is very challenging due to mathematical intractability, various throughput--hit-rate tradeoff designs obtained by considering tractable random-push scheduling can serve as alternatives. This is because priority-push and random-push scheduling have similar behaviors, except that the former scheduling prioritizes the users that can use D2D links.
\end{itemize}
}

\vspace{-15pt}

\section{Conclusions}
In this work, we used the individual preferences of users to improve cache-aided D2D networks. We used an individual preference probability model to derive the network utility of a clustering network and to propose a utility maximization problem accordingly. This problem can be applied to solve different important and practical problems, e.g., throughput, EE, hit-rate optimization, and different tradeoff problems. Assuming users can coordinate, we proposed a general solution approach for solving the utility maximization problem. Comprehensive numerical evaluations were conducted with practical individual preferences and network setups. Our results show that we can appropriately exploit information about users' individual preferences to significantly increase the performance of cache-aided D2D networks. Our results also show that throughput and hit-rate are in conflict with each other; nevertheless, such a conflict can be resolved through a suitable tradeoff design. To obtain an effective EE design, in addition to directly optimizing EE, we can solve a properly designed throughput--hit-rate tradeoff design, offering another perspective for EE optimization. Aside from optimizing the caching policy to improve performance, we proposed changing the cooperation distance of the D2D to achieve this goal; likewise, the tradeoff exists in this regard. Finally, we demonstrated that the results of our work can serve as a foundation for designing caching policies in networks with a more involved scheduling policy.

\appendices
\section{Derivations of the Expected Utility}
We first derive the expression of $U$. Using (\ref{eq:BS_rate}), (\ref{eq:self_rate}), and (\ref{eq:D2D_rate}), we obtain
\begin{equation}
\begin{aligned}
\nonumber
 U&=\sum_{k\in\mathcal{U}_A} \frac{w_k}{K_{\text{A}}}\mathbb{E}\Bigg\lbrace U_{\text{D}}\left(1-\sum_{m=1}^M a_m^k\left[\prod_{l\in \mathcal{U}}(1-b_m^l\mathbf{1}_{\lbrace h_{k,l},C\rbrace})\right]-\sum_{m=1}^M a_m^kb_m^k\right)\\
&\qquad +U_{\text{B}}\sum_{m=1}^M a_m^k\left[\prod_{l\in \mathcal{U}}(1-b_m^l\mathbf{1}_{\lbrace h_{k,l},C\rbrace})\right] + U_{\text{S}}\sum_{m=1}^M a_m^kb_m^k \Bigg\rbrace\\
\end{aligned}
\end{equation}
\begin{equation}
\begin{aligned}
\label{eq:App_reformulate}
&=\sum_{k\in\mathcal{U}_A} \frac{w_k U_{\text{D}}}{K_{\text{A}}}+(U_{\text{B}}-U_{\text{D}})\sum_{k\in\mathcal{U}_A}\sum_{m=1}^M \frac{w_ka_m^k}{K_{\text{A}}}\left[\prod_{l\in\mathcal{U}}(1-b_m^lE_{h}\lbrace\mathbf{1}_{\lbrace h_{k,l},C\rbrace}\rbrace)\right]+(U_{\text{S}}-U_{\text{D}})\sum_{m=1}^M \sum_{k\in\mathcal{U}_A} \frac{w_k a_m^kb_m^k}{K_{\text{A}}}\\
&=\sum_{k\in\mathcal{U}_A} \frac{w_k U_{\text{D}}}{K_{\text{A}}}+(U_{\text{B}}-U_{\text{D}})\sum_{m=1}^M\underbrace{\sum_{k\in\mathcal{U}_A} \frac{w_ka_m^k}{K_{\text{A}}}\left[\prod_{l\in\mathcal{U}}(1-b_m^lL_{k,l})\right]}_{S_m}+(U_{\text{S}}-U_{\text{D}})\sum_{m=1}^M \sum_{k\in\mathcal{U}_A} \frac{w_ka_m^kb_m^k}{K_{\text{A}}}\\
&=\sum_{k\in\mathcal{U}_A} \frac{w_k U_{\text{D}}}{K_{\text{A}}}+(U_{\text{B}}-U_{\text{D}})\sum_{m=1}^M S_m+(U_{\text{S}}-U_{\text{D}})\sum_{m=1}^M \sum_{k\in\mathcal{U}_A} \frac{w_ka_m^kb_m^k}{K_{\text{A}}}.
\end{aligned}
\end{equation}
By using (\ref{eq:App_reformulate}), we thus obtain
\begin{equation}
\begin{aligned}\nonumber
U_{\text{net}} &=\sum_{k\in\mathcal{U}_A} \frac{w_k U_{\text{D}}}{K_{\text{A}}}+(U_{\text{B}}-U_{\text{D}})\sum_{m=1}^M S_m+(U_{\text{S}}-U_{\text{D}})\sum_{m=1}^M \sum_{k\in\mathcal{U}_A}\frac{w_ka_m^kb_m^k}{K_{\text{A}}} + U_{\text{S}}\sum_{m=1}^{M}\sum_{k\in\mathcal{U}_A}\sum_{l\in\mathcal{U}_A,l\neq k} \frac{w_la_m^lb_m^l}{K_{\text{A}}}\\
&=\sum_{k\in\mathcal{U}_A} \frac{w_k U_{\text{D}}}{K_{\text{A}}}+(U_{\text{B}}-U_{\text{D}})\sum_{m=1}^M S_m +(K_{\text{A}}U_{\text{S}}-U_{\text{D}})\sum_{m=1}^M\sum_{k\in\mathcal{U}_A} \frac{w_ka_m^kb_m^k}{K_{\text{A}}}.
\end{aligned}
\end{equation}

\section{Proof of Theorem 1}
To prove the Theorem, we first note that the problem in (\ref{eq:optmization_Ut}) satisfies the block separable structure as follows:
\begin{equation}
\begin{array}{cl}\label{eq:Problem_BCD}
\displaystyle{\max_{\mathbf{b}_1,...,\mathbf{b}_K}} &U(\mathbf{b}_1,\mathbf{b}_2,...,\mathbf{b}_K)\\
\text{s.t.} &\mathbf{b}_k\in\mathcal{B}_k,\forall k.
\end{array}
\end{equation}
Eq. (\ref{eq:Problem_BCD}) indicates that the constraints on different blocks are separable. Denote $u(\mathbf{b}_{k'};\mathbf{B}^r)=U(\mathbf{b}_1^{r},...,\mathbf{b}_{k'-1}^{r},\mathbf{b}_{k'},\mathbf{b}_{k'+1}^{r},...,\mathbf{b}_{K}^{r})$ for brevity. From Alg. \ref{alg:BCD}, we notice that $\mathbf{b}_{k'}^{r+1}=\displaystyle{\arg\max_{\mathbf{b}_{k'}\in\mathcal{B}_i}}\text{ } u(\mathbf{b}_{k'};\mathbf{B}^r)$ at each iteration. Hence, we have
\begin{equation}
u(\mathbf{b}_{k'}^{r+1};\mathbf{B}^r) \geq u(\mathbf{b}_{k'}^r;\mathbf{B}^r).
\end{equation}
Thus, we know that the algorithm is monotonically non-decreasing. Then, since the optimal objective function of (\ref{eq:optmization_Ut}) should not be infinity, the algorithm must converge.

To prove that Alg. \ref{alg:BCD} converges to a stationary point if every iteration has an unique maximization, we use the analysis framework for block coordinate descent methods in \cite{Bertsekas:NP} as follows.\footnote{The proof basically follows the same steps as those taken to prove Proposition 2.7.1 in \cite{Bertsekas:NP}. Hence, we here provide a shortened version of the proof.} Suppose each iteration in Alg. \ref{alg:BCD} has a unique maximizer. Then, Alg. \ref{alg:BCD} converges to a unique solution $\overline{\mathbf{B}}=\left[\bar{\mathbf{b}}_1,...,\bar{\mathbf{b}}_K\right]$ as the number of iterations $r\to\infty$. According to Alg. \ref{alg:BCD}, we know that
\begin{equation}
\bar{\mathbf{b}}_{k'} = \displaystyle{\arg\max_{\mathbf{b}_{k'}\in\mathcal{B}_{k'}}}\text{ } u(\mathbf{b}_{k'};\overline{\mathbf{B}}),\forall k'=1,2,...,K.
\end{equation}
As a result, due to concavity,
\begin{equation}\label{Indi_App_Thm1}
\left(\nabla_{k'} U(\bar{\mathbf{b}}_1,...,\bar{\mathbf{b}}_K)\right)^T\left(\mathbf{b}_{k'}-\bar{\mathbf{b}}_{k'}\right)\leq 0,\forall \mathbf{b}_{k'}\in\mathcal{B}_{k'},
\end{equation}
where $\nabla_{k'}$ denotes the gradient of $U$ with respect to component $\mathbf{b}_{k'}$. We denote $\mathbf{b}=vec(\mathbf{B})\in \mathcal{B}$ is the vectorization of $\mathbf{B}=\left[\mathbf{b}_1,..,\mathbf{b}_K\right]$ and $\bar{\mathbf{b}}=vec(\overline{\mathbf{B}})\in \mathcal{B}$ is the vectorization of $\overline{\mathbf{B}}$, where $\mathcal{B}=vec(\mathcal{B}_1 \times \mathcal{B}_2\times ...\times \mathcal{B}_K)$. Then, notice that (\ref{Indi_App_Thm1}) is true for all $k'=1,2,...,K$. It follows from the Cartesian product structure of a set that
\begin{equation}
\left(\nabla U(\bar{\mathbf{b}})\right)^T\left(\mathbf{b}-\bar{\mathbf{b}}\right)\leq 0,\forall \mathbf{b}\in\mathcal{B},
\end{equation}
where $\nabla$ is the gradient with respect to $\bar{\mathbf{b}}$. This proves that Alg. \ref{alg:BCD} converges to a stationary point.



%\bibliographystyle{IEEEtran}

\end{document}